\documentclass[aps,prb,reprint,groupedaddress,floatfix]{revtex4-1}

\usepackage[english]{babel} 
\usepackage{amssymb,amsmath}
\usepackage{graphicx}
\usepackage{url}
\usepackage{multirow}
\usepackage{chemformula}
\usepackage{tabularx}
\usepackage{makecell}
\usepackage{xcolor}
\usepackage{siunitx}

\usepackage{subcaption}

\newlength{\smallpic}
\begin{document}

\title{Effect of Hubbard U corrections on the electronic and magnetic properties of 2D materials: A high-throughput study}
\author{Sahar Pakdel$^*$, Thomas Olsen, and Kristian S. Thygesen$^{**}$ 
 }
 \address{Computational Atomic-scale Materials Design (CAMD), Department of Physics, Technical University of Denmark, 2800 Kgs. Lyngby Denmark \\ \normalfont{$^*$Corresponding author: sahpa@dtu.dk}
\normalfont{$^{**}$thygesen@fysik.dtu.dk}}
\date{\today}

\begin{abstract}
We conduct a systematic investigation of the role of Hubbard U corrections in electronic structure calculations of two-dimensional (2D) materials containing $3d$ transition metals. Specifically, we use density functional theory (DFT) with the PBE and PBE+U approximations to calculate the crystal structure, band gaps, and magnetic parameters of 638 monolayers. Based on a comprehensive comparison to experiments we first establish that inclusion of the U correction worsens the accuracy for the lattice constant. Consequently, PBE structures are used for subsequent property evaluations. The band gaps show significant dependence on the U-parameter. In particular, for 134 (21\%) of the materials the U parameter leads to a metal-insulator transition. For the magnetic materials we calculate the magnetic moment, magnetic exchange coupling, and magnetic anisotropy parameters. In contrast to the band gaps, the size of the magnetic moments shows only weak dependence on U. Both the exchange energies and magnetic anisotropy parameters are systematically reduced by the U correction. On this basis we conclude that the Hubbard U correction will lead to lower predicted Curie temperatures in 2D materials. All the calculated properties are available in the Computational 2D Materials Database (C2DB).

\end{abstract}

\maketitle

\section{Introduction}
 Solving the electronic structure problem for a periodic solid with localised electrons represents a long standing challenge in condensed matter physics dating back to the seminal work of Hubbard\cite{hubbard1964electron}. Within the paradigmatic Hubbard model the degree of localisation can be quantified by the dimensionless parameter $U/t$, where $U$ is the on-site screened Coulomb interaction energy and $t$ is the hopping energy driving the delocalization of states. The most interesting and illustrative regime of the Hubbard model is obtained for a half filled band when $U$ is much larger than $t$. In this case, the electrons will tend to localise on the atomic sites in order to minimise the Coulomb repulsion. However, even in the limit of $U\gg t$, the electrons will not fully localise and the ground state will contain a small fraction of doubly occupied sites. These components of the wave function are necessary to activate the hopping term and leads to a small energy gain relative to the fully localised states (due to spin there are $2^N$ such states each with energy $N\varepsilon_0$). Although the exact solution of the two-dimensional (2D) Hubbard model is not known, its ground state is believed to be a strongly correlated paramagnetic state with some local anti-ferromagnetic order\cite{mancini2000mott}. In this (almost) fully localised state, there will be an energy cost of approximately $U+\varepsilon_0$ associated with the addition of an electron to the system. On the other hand, removing an electron costs only $\varepsilon_0$. The band gap is thus equal to $U$. The existence of such a band gap is the hallmark of the correlated Mott insulating phase\cite{mott1968metal}. 
 
 While the true ground state of the Hubbard model is most likely paramagnetic (with local ferromagnetic order), systems in the Mott regime may spontaneously break time-reversal symmetry and exhibit long range magnetic order at low temperature. In contrast to the correlated Mott state, such symmetry broken states may be well described within a mean field approximation to the Hubbard model (the Stoner model of metallic ferromagnetism\cite{stoner1938collective}). In the mean field treatment, the magnetic state is stabilised by the exchange mechanism. While the fully polarised mean field solution is very different from the true correlated ground state, the total energy gap between them becomes small when $t\ll U$ and some key properties of the states become similar (the total magnetic moment is clearly not one of them). In particular, both states describe localised electrons and exhibit a quasiparticle gap of the order of $U$. These similarities are the main reason for the success of the DFT+U method to be explored in the present work.  

 Among the computational methods capable of describing strongly correlated solids like Mott insulators, are dynamical mean field theory\cite{metzner1989correlated,georges1992hubbard,georges1996dynamical} and reduced density matrix functional theory\cite{yang2000degenerate}. Unfortunately, these approaches are computationally demanding and come with their own challenges and limitations. We note that the random phase approximation (RPA) has been advocated as an accurate, fully \emph{ab initio}, and computationally affordable beyond-DFT total energy method. Indeed, the RPA provides an excellent description of long-range correlations, including van der Waals interactions, and yields a decent description of formation energies and structural parameters of simple solids\cite{harl2010assessing,olsen2013random,jauho2015improved}. However, its failure to capture short-range correlations, makes it unsuitable for systems with localised states. In fact, it is unclear whether any diagrammatic expansion (beyond RPA) around a non-interacting ground state\cite{olsen2012extending,gruneis2009making,olsen2014accurate,olsen2019beyond,patrick2015adiabatic} can yield a valid and sufficiently accurate description of the most complex such systems\cite{schafer2016nonperturbative}.     

Density functional theory (DFT)\cite{hohenberg1964inhomogeneous,kohn1965self} is by far the most popular method for solving the electronic structure problem thanks to its optimal balance between accuracy and computational cost. However, when it comes to solids with localised valence states (typically, $d$ or $f$ orbitals), in particular Mott insulators, DFT presents rather severe and well documented shortcomings. Indeed, electron localisation enhances the self-interaction and delocalisation errors of the widely used (semi)local density functional approximations\cite{cohen2008insights,perdew1981self} resulting in qualitatively wrong ground state properties, e.g. incorrect magnetic moments/ordering or even vanishing band gaps in (Mott) insulators. In a way such failures should come as no surprise since DFT (within the Kohn-Sham scheme) is an effective single-particle theory while the Mott insulating state is strongly correlated. However, it turns out that some of the key properties of solids with localised electrons, in particular the magnetic properties and finite band gap, may be correctly and efficiently predicted by introducing a Hubbard U correction to the Kohn-Sham Hamiltonian\cite{anisimov1991density,anisimov1991band,anisimov1993density,solovyev1994corrected,anisimov1997first,liechtenstein1995density}. \textcolor{black}{The Coulomb parameter U in such approaches can either be chosen from empirical studies\cite{PhysRevB.73.195107}, from linear response calculations\cite{PhysRevB.71.035105} or from direct evaluation of Coulomb integrals in a localized basis\cite{PhysRevMaterials.5.034001}. While the latter two methods both entail first principles methods for obtaining U the physical contents are rather different. The linear response method aims at correcting inherent errors introduced by common exchange-correlation functionals, while the Coulomb integrals represent actual parameters to be included in a Hubbard model description of a given material.}

The use of such DFT+U methods has been thoroughly explored for bulk solids with partially filled $d$ or $f$ shells\cite{himmetoglu2014hubbard,vaugier2012hubbard,orhan2020first,timrov2018hubbard,shenton2017effects,tancogne2018ultrafast,agapito2015reformulation}. However, much less is known about the role of Hubbard U corrections in DFT calculations for the emergent class of atomically thin two-dimensional (2D) materials\cite{huang2020first,pasquier2022ab,das2019band,li2020high}. The physics of localised electrons in such systems can be expected to differ from that of the traditional bulk solids for two main reasons. First, the reduced coordination number of atoms hosting the localised states should lead to narrower bands. Secondly, as dielectric screening is weaker in two-dimensional electron systems \cite{cudazzo2011dielectric,huser2013dielectric}, the (screened) Coulomb interaction should be larger. From these considerations one would expect localisation-driven correlation effects to be even more important in 2D than in 3D solids.   

In this work we perform a systematic investigation of Hubbard U-corrections in DFT calculations for 2D materials. Specifically, we calculate lattice constants, electronic band gaps, magnetic moments, magnetic exchange coupling, and magnetic anisotropy parameters for a set of 638 2D materials all containing one or more $3d$ transition metals. Throughout this work, the Hubbard U correction is applied to the $3d$ orbitals of the transition metal atoms only. For a small set of materials we first explore how the various observables depend on the value of U. Based on this pre-study, we set the value of U to 4 eV for all the transition metals. \textcolor{black}{This value coincides with the value used for most of the $3d$ transition metal oxides in the OQMD.\cite{Saal2013}} The application of the U-correction is found to worsen the agreement with experiments for the lattice constants and therefore PBE relaxed structures are considered throughout. We find that U-corrections have only negligible effect on the size of the magnetic moments and whether or not the material is predicted to be magnetic. Band gaps are generally increased by the U-correction and no less than 134 (21\%) of the materials undergo a metal-insulator transition. The exchange couplings are significantly reduced by the U-correction, which we ascribe to the stronger localisation of the states involving the $3d$ orbitals. Finally, the Hubbard correction yields a significant decrease of magnetic anisotropy energies, which may be ascribed to a reduction of crystal field effects due to increased orbital localization.

\begin{figure*}
    \centering
    \includegraphics[width=\textwidth]{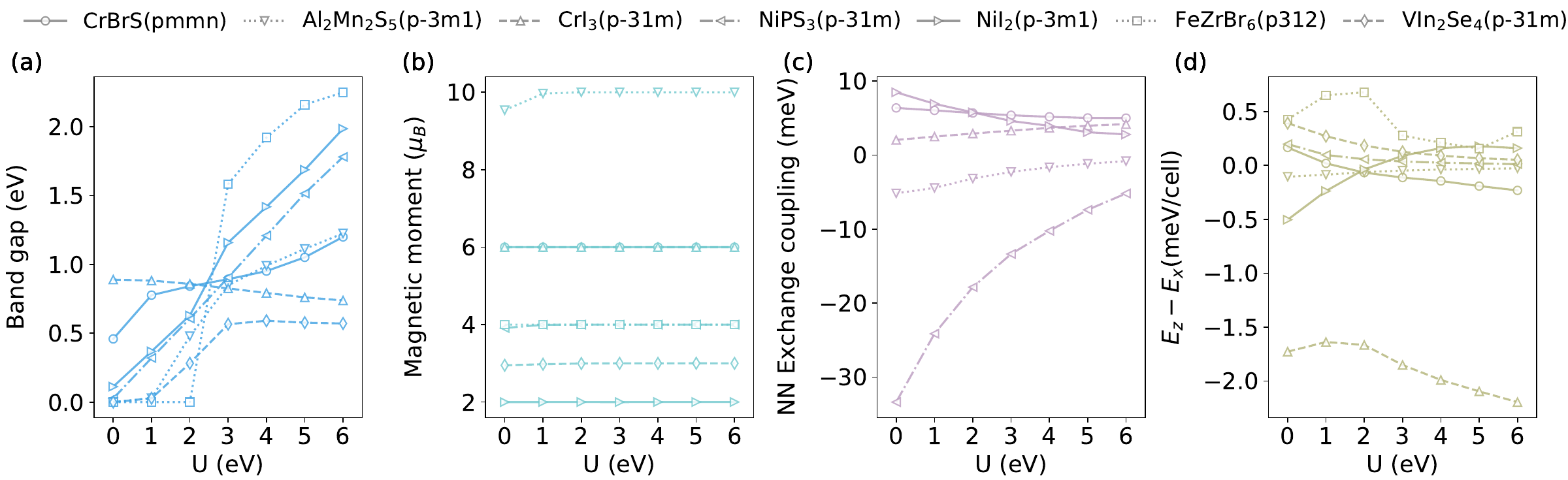}
    \caption{The dependence of different properties on the Hubbard U-correction for seven representative monolayers (listed by their chemical formula and layer group). The calculated properties are: (a) The electronic band gap. (b) The total magnetic moment of the unit cell. (c) The nearest neighbor exchange coupling. (d) The magnetic anisotropy. }
    \label{fig:Udependence}
\end{figure*}

We stress that the purpose of this work is not to make a detailed assessment of the accuracy of the DFT+U method by comparing to higher level theories or experiments (with the exception of in-plane lattice constants where abundant and reliable experimental data is available). Nor is the purpose to determine the most accurate value of the U parameter for specific elements/materials. \emph{Rather, it is our objective is to present a consistent, systematic, and unbiased comparison of PBE and PBE+U for a range of different properties across many different 2D materials.}

All the results are available as part of the Computational 2D Materials Database (C2DB)\cite{haastrup2018computational,gjerding2021recent}.

\section{Methods}
The computational workflow was constructed within the Atomic Simulation Recipes (ASR) Python framework\cite{gjerding2021atomic} and executed using the MyQueue\cite{mortensen2020myqueue} job scheduler.
All calculations were performed with the GPAW electronic structure code\cite{mortensen2024gpaw,enkovaara2010electronic} using a plane wave cut off of 800 eV and a $k$-point density of 12 \AA. For DFT+U calculations GPAW employs an effective U-parameter to account for on-site Coulomb and exchange interactions,
\begin{equation}
    U^{\mathrm{eff}}=U-J.
\end{equation}
The total energy then takes the form  
\begin{equation}\label{eq:Hubbard}
    E_{\mathrm{DFT+U}}=    E_{\mathrm{DFT}}+\sum_{a,i}\frac{1}{2} U_i^{\mathrm{eff}}\mathrm{Tr}(\rho_i^a-\rho_i^a\rho_i^a)
\end{equation}
where the sum runs over atoms ($a$) and the atomic orbitals ($i$) for which the correction is applied (the 3d orbitals in the present work). $\rho_i^a$ is the density matrix in the basis of the orbitals $i$ (only the 3d orbitals, so the sum over $i$ is obsolete). We only included the Hubbard corrections to the elements V, Cr, Mn, Fe, Co, Ni, Cu, which have previously 
been identified as the most important elements \cite{PhysRevB.73.195107} and the we set $U_i^{\mathrm{eff}}$ to 4 eV for the $3d$ orbitals.  All structures considered in this work have been relaxed until the maximum force on the atoms is below 0.01 eV/\AA.

\section{The Materials}
The materials considered are monolayers from the C2DB\cite{haastrup2018computational} containing at least one of the seven $3d$ transition metals V, Cr, Mn, Fe, Co, Ni, Cu. In addition, we require that the monolayer should have an energy above the convex hull below 0.2 eV/atom (when evaluated with PBE). This requirement ensures that all the materials are chemically "reasonable" such that conclusions drawn from the data set are not influenced by the presence of unrealistic structures or chemistry. Finally, we only calculate properties for monolayers where the PBE+U yields a finite band gap. This is done because the Hubbard U approach can only be physically justified for non-metallic systems.    

\begin{figure*}
    \centering
    \includegraphics[width=\textwidth]{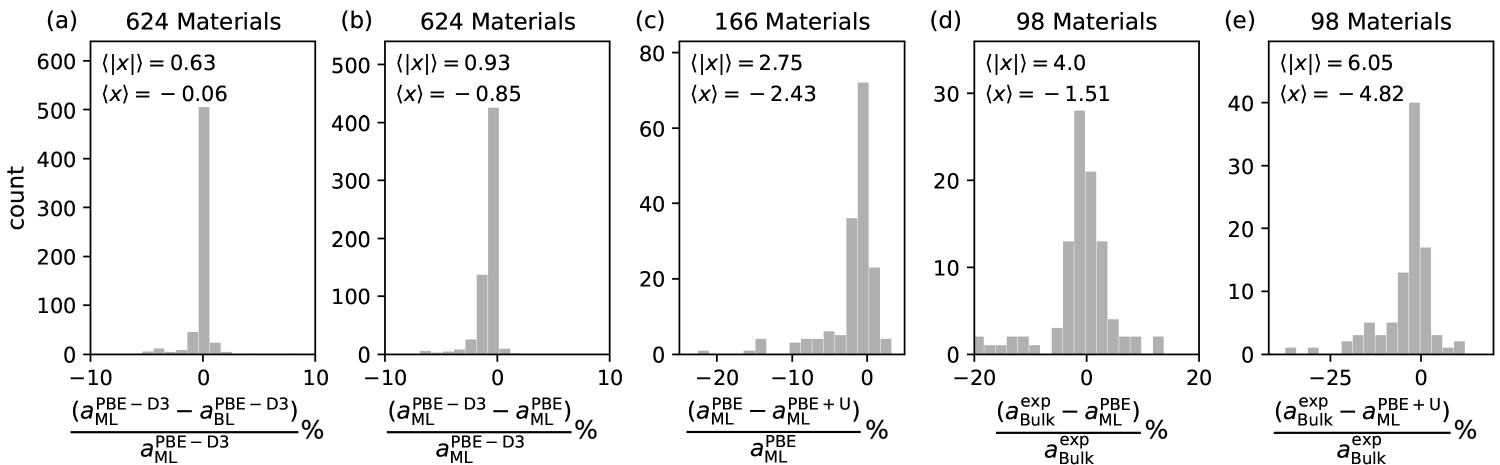}
    \caption{In-plane lattice constants ($a$) of monolayers, bilayers, and bulk. (a) The difference in $a$ for monolayer and homobilayer structures evaluated with the PBE-D3 functional. (b) The difference in $a$ for monolayers calculated with the PBE and PBE-D3 functionals, respectively. (c) The difference in $a$ for 166 monolayers calculated with PBE and PBE+U (U=4 eV). (d) The difference in $a$ for 98 materials in bulk form (experimental) and monolayer form (PBE). (e) The difference in $a$ for 98 materials in bulk form (experimental) and monolayer form (PBE+U). }
    \label{fig:lattice}
\end{figure*}

\section{Results}
Before analysing the effect of the Hubbard U on different properties across the full set of materials, we explore how each of the properties depend on U for a small set of representative materials. Based on this analysis we set the value of the U-correction to 4 eV in all subsequent calculations. This choice is made based on the principle of Occam's razor. Subsequently, we show that inclusion of U worsens the agreement with experiments for the in-plane lattice constants of the monolayers. Consequently, all subsequent property calculations are performed for the PBE relaxed monolayers. Finally, we analyse the effect of U on the band gap and magnetic properties across all the monolayers.   

\subsection{U-dependence}
Figure \ref{fig:Udependence} shows the U-dependence of the four properties investigated in this work. The chemical formula and the layer group of the materials are listed at the top of the figure. In all cases the monolayers have been relaxed with the PBE xc-functional (see next section). We stress that the seven materials have been selected to illustrate different types of U-behavior and thus do not necessarily comprise a representative subset of the full set of materials explored later in the paper. 

Panel (a) shows the U-dependence of the electronic band gap. In general, the band gap shows a strong variation with U. The typical behavior is that the band gaps increase monotonically with U, but there are also deviations from this trend. The band gap of CrI$_3$ decreases slightly with U. The band gap of VIn$_2$Se$_4$ first increases with U, but then saturates and remains constant form $U>3$eV. The band gap of FeZrBr$_6$ remains zero until U=2 eV and then increases sharply.  
In general, it is difficult to predict or even rationalise the U-dependence of the band gap for a given material. 

In contrast to the band gap, the magnetic moment shown in panel (b) does not show any significant dependence on U. In fact, the magnetic moment of all the materials are completely independent of U. The weak (vanishing) U-dependence of the magnetic moment follows from the fact that for an insulator, the magnetic moment must be an integer. Thus changing the magnetic moment is not possible without going through a metal transition. As an example of the latter case, we mention Al$_2$Mn$_2$S$_5$, which is metallic for $U=0$ with a non-integer magnetic moment. For finite U a gap opens and the magnetic moment takes an integer value which stays constant for all larger values of U. 

The nearest neighbor exchange coupling, $J$, shown in panel (c) generally decreases with U. This behavior can be understood from the general tendency of the U-correction to localise the $d$-electrons and thereby reduce the exchange integrals of the $d$-states. Since the magnetism is mainly carried by the $d$-electrons, this explains the observed trend. 

\begin{figure*}
    \centering
    \includegraphics[width=0.9\textwidth]{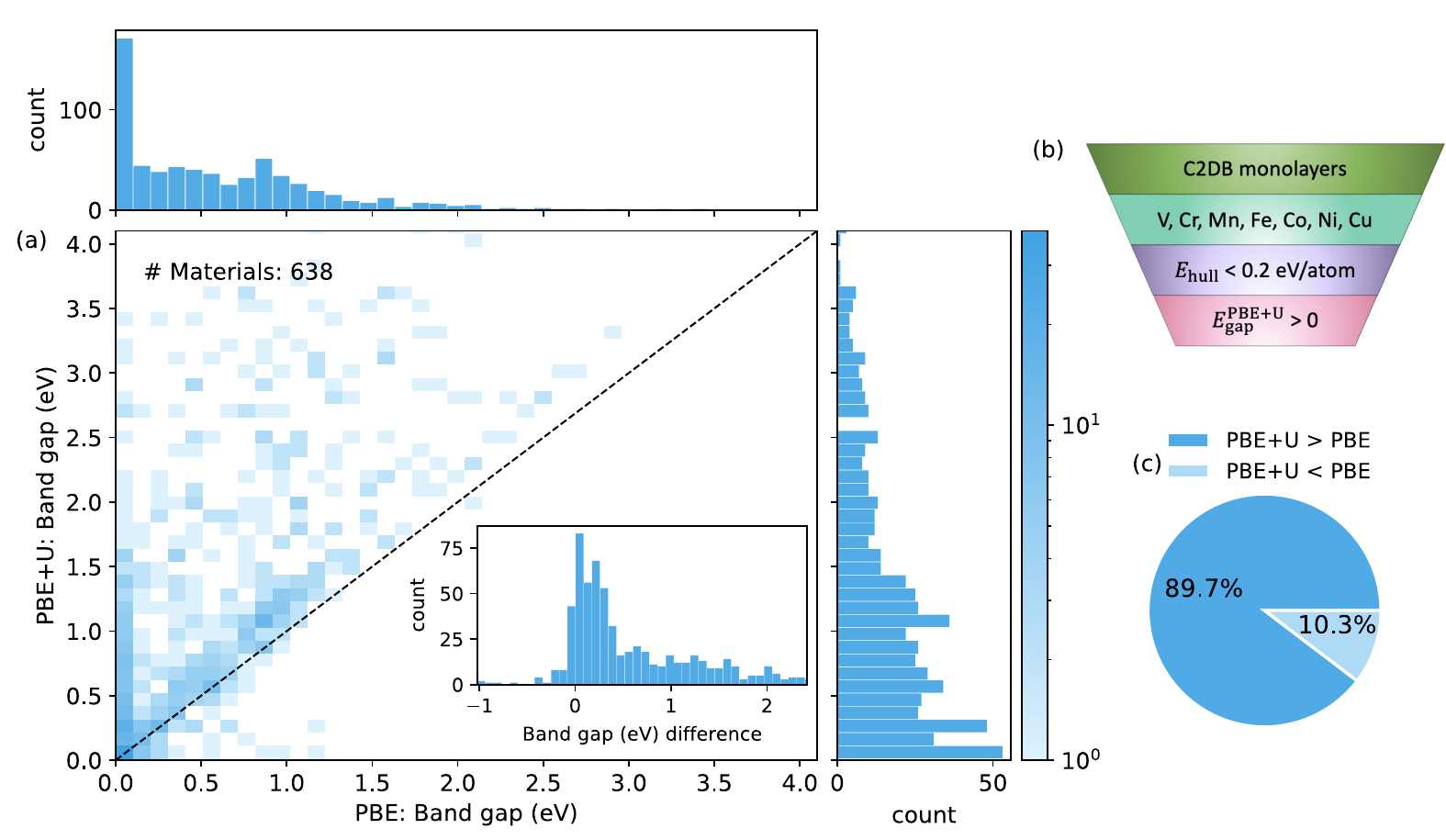}
    \caption{(a) Band gaps calculated with PBE versus PBE+U for 638 monolayers. The side panels shows the distribution of band gaps obtained with the two methods, and the inset shows distribution of the the band gap \emph{difference}. The screening criteria used to select the materials from the C2DB database are shown in panel (b). The pie chart in panel (c) shows the fraction of materials for which the band gap is increased/decreased by the Hubbard-U correction. All PBE+U calculations were performed with a Hubbard-correction of 4 eV. }
    \label{fig:gaps}
\end{figure*}

The magnetic anisotropy (MA) is defined as the difference in total energy between the out-of-plane direction ($z$) and in-plane direction ($x$) of the magnetic polarisation\cite{torelli2018calculating}. It is thus a spin-orbit effect, which explains its small value on the order of 1 meV. A negative value corresponds to an out-of-plane magnetic easy axis while a positive value signals a magnetic easy plane. Like the exchange coupling, the MA shows a decreasing trend with U. Again, this can be explained by the increased localisation of the $d$-orbitals. Indeed, stronger localisation will bring the shape of the $d$-states closer to the isotropic atomic orbitals (the MA vanishes for an isolated atom). Only CrI$_3$ deviates from this trend as its MA increases (becomes more negative) with U. This deviation can be rationalized from the fact that the nearest neighbor exchange constant in CrI$_3$ increases with U and so does the anisotropic exchange, which dominates the magnetic anisotropy in the PBE+U description\cite{Olsen2019}.

As we shall see in the following sections, many of the trends observed above for the seven monolayers carry over to the full set of monolayers. For simplicity, and because no significant qualitative changes occur as function of U, we shall from hereon fix the U-parameter to 4 eV for all the relevant elements (V, Cr, Mn, Fe, Co, Ni, Cu). The results obtained for U=4 eV are available in the C2DB together with the PBE results. In practical situations one can get a feeling for the effect of U by comparing results for U=0 eV (pure PBE) and U=4 eV.

\subsection{Lattice constants}
To assess the accuracy of the PBE+U for the (in-plane) lattice constants of monolayers we compare to an experimental data set consisting of the in-plane lattice constant, $a$, of 98 layered van der Waals (vdW) crystals. These crystals all represent the bulk form of one of the monolayers considered in this work. For in-plane anisotropic materials we select the largest lattice constant. 

\begin{figure*}
    \centering
    \includegraphics[width=0.9\textwidth]{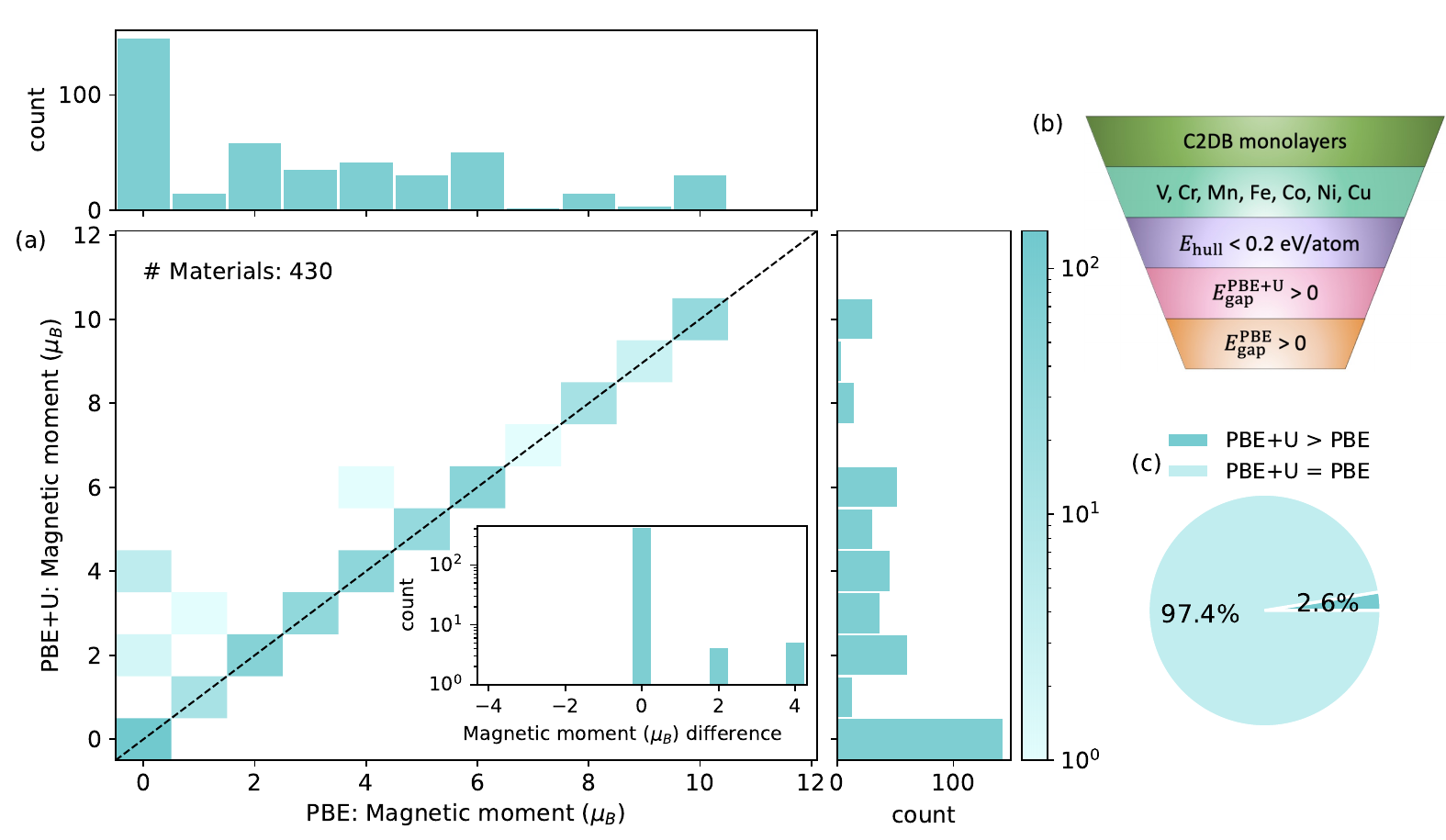}
    \caption{(a) Total magnetic moment per unit cell calculated with PBE versus PBE+U for 430 monolayers. Only ferromagnetic spin configurations have been considered. The side panels shows the distribution of the magnetic moments obtained with the two methods, and the inset shows their \emph{difference}. The screening criteria used to select the materials from the C2DB database are shown in panel (b). The pie chart in panel (c) shows the fraction of materials for which the magnetic moment is unchanged/increased by the Hubbard-U correction. All PBE+U calculations were performed with a Hubbard-correction of 4 eV.}
    \label{fig:magmom}
\end{figure*}

\begin{figure*}
    \centering
    \includegraphics[width=0.9\textwidth]{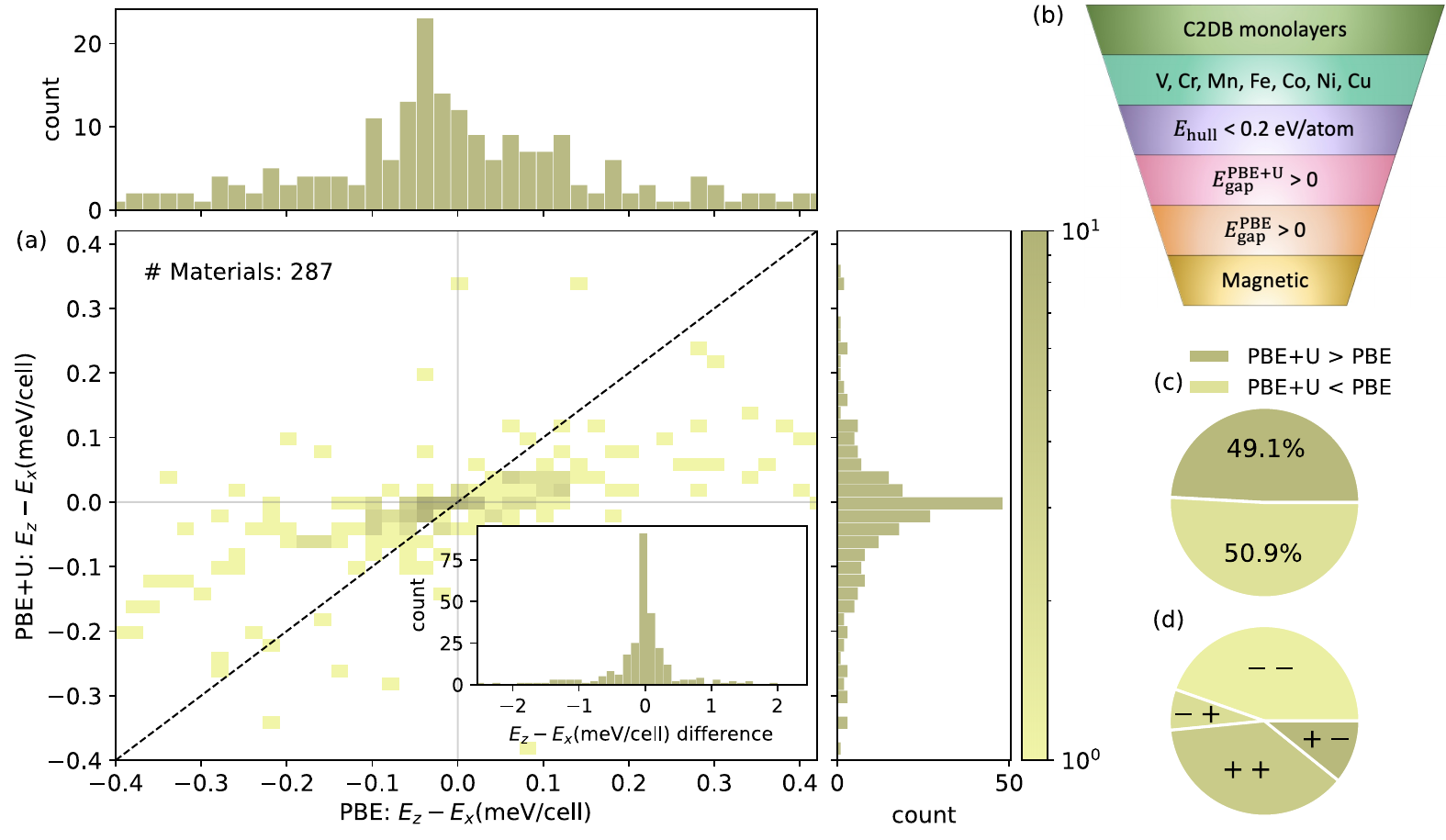}
    \caption{(a) Magnetic anisotropy calculated with PBE versus PBE+U for 287 monolayers. The side panels shows the distribution of the magnetic anisotropy obtained with the two methods, and the inset shows their \emph{difference}. The screening criteria used to select the materials from the C2DB database are shown in panel (b). The pie chart in panel (c) shows the fraction of materials for which the absolute size of the magnetic anisotropy is increased/decreased by the Hubbard-U correction. The pie chart in panel (d) shows how the distribution of materials according to the sign of the magnetic anisotropy predicted by the two methods. All PBE+U calculations were performed with a Hubbard-correction of 4 eV.}
    \label{fig:anisotropy}
\end{figure*}

Figure \ref{fig:lattice} summarises the results of our lattice constant analysis. In all calculations the structure and unit cells have been fully relaxed. The first question concerns the justification for comparing in-plane lattice constants of the bulk and monolayers. Since all the monolayers included in the study have high thermodynamic stability (all materials have formation energy within 0.2 eV/atom of the convex hull\cite{gjerding2021recent}), they are not expected to form chemical bonds upon stacking. Consequently, the interlayer interactions in the bulk should be of the weak vdW type and thus not influence the structure of the individual monolayers much. Indeed, our PBE-D3 calculations including vdW corrections, show that the in-plane lattice constants of bilayers deviate only little from those of the monolayers with a mean absolute relative deviation (MARD) of only 0.63\%, see Figure \ref{fig:lattice}(a). This justifies the comparison of calculated lattice constants of monolayers and experimental lattice constants of bulk. The bilayer structures used for the comparison in \ref{fig:lattice}(a) represent the most stable stacking configuration contained in the Computational Bilayer Database (BiDB)\cite{pakdel2023emergent}. 

Figure \ref{fig:lattice}(b) shows that the PBE and PBE-D3 yields very similar lattice constants for the monolayers. As expected the inclusion of the attractive D3 vdW-correction systematically reduces the lattice constant, but only by 0.93\% on average. This result in not used directly in the current work, but is included here for completeness.

Figure \ref{fig:lattice}(c) compares the lattice constants obtained with PBE and PBE+U (with U=4 eV). It is clear that the U-correction systematically expands the lattice. This can be understood from the physics described by the Hubbard-term, see Eq. (\ref{eq:Hubbard}) Indeed, the Hubbard term penalises partial occupations of the $d$-orbitals and thus drives the system towards integer occupations. This is energetically easier to achieve for more narrow $d$-bands. Consequently, the Hubbard-term will tend to increase the lattice constants in order to reduce hybridisation and enhance the localisation. We stress that there are materials that deviate from this trend. \color{black}Finally, we mention that we have also studied whether the trend of worsened lattice constant with U can be ascribed to a particular family of materials. Specifically, we have considered the subsets of halides and chalcogenides separately. For these subsets, however, we found the same trend as for the entire set of materials.\color{black}

Finally, Figure \ref{fig:lattice}(d,e) shows the comparison between the lattice constants of the monolayers calculated with PBE and PBE+U, respectively, and the experimental lattice constants of the parent bulk crystals. It can be seen that PBE yields a lower MARD than PBE+U (4.0\% versus 6.05\%). In addition, PBE shows a significantly smaller systematic error as compared to PBE+U, which systematically overestimates the lattice constant. We note that the deviations found here are larger than expected for the PBE and PBE+U functionals. Some of this deviation may be ascribed to the fact that we compare monolayers to bulk, although panel (a) suggests this to be a small effect. Thus the main explanation for the larger than expected deviation is that the materials considered all contain $3d$ elements that are inherently difficult to describe. In fact, a similar comparison performed for 164 monolayers not containing $3d$ metals yielded a MARD of only 1.49\% (see SI of Ref. \cite{pakdel2023emergent}).

\begin{figure*}
    \centering
    \includegraphics[width=0.9\textwidth]{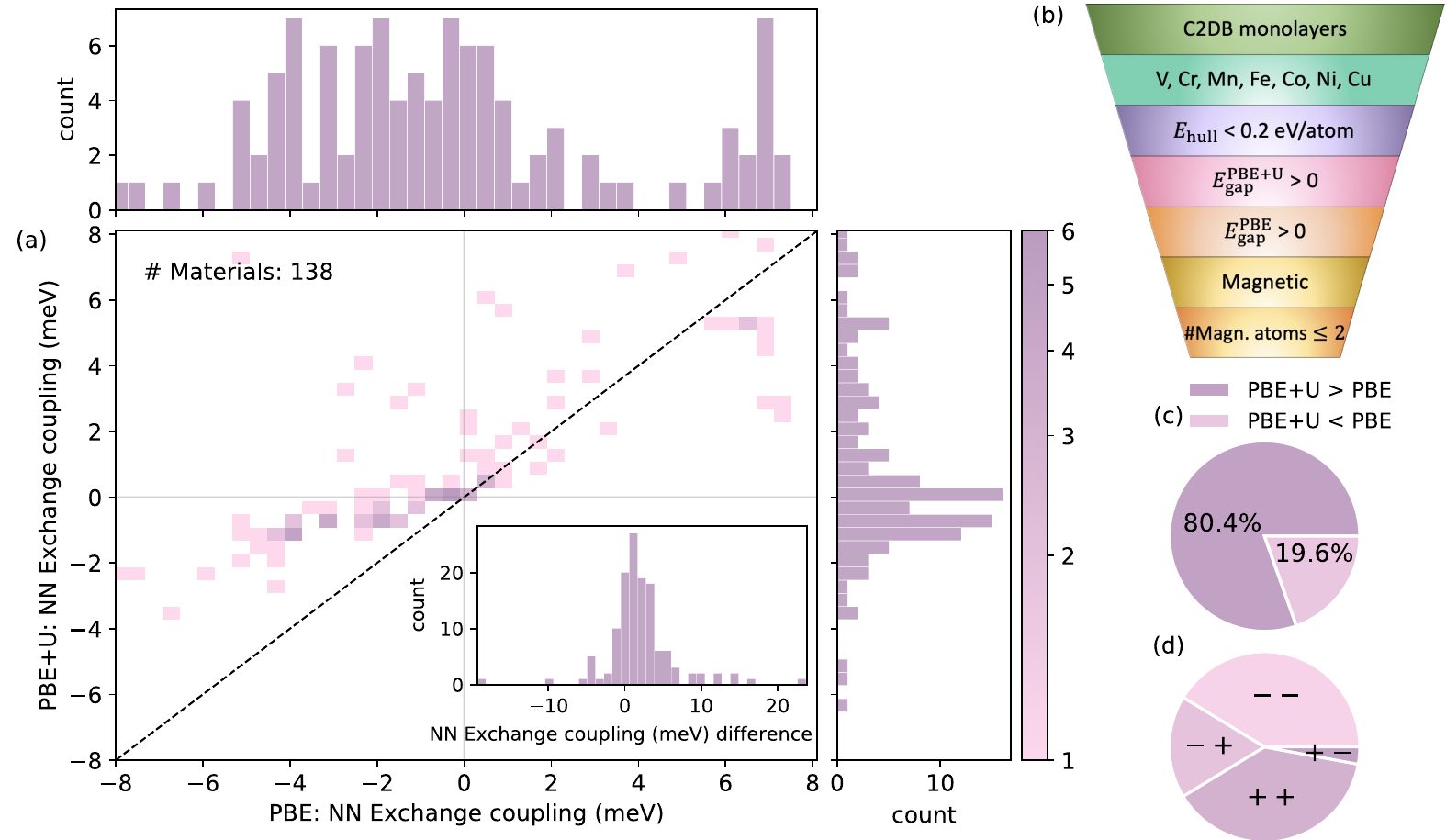}
    \caption{(a) Magnetic exchange coupling per magnetic element (V, Cr, Mn, Fe, Co, Ni, Cu) calculated with PBE versus PBE+U for 138 monolayers. The side panels shows the distribution of the exchange coupling obtained with the two methods, and the inset shows their \emph{difference}. The screening criteria used to select the materials from the C2DB database are shown in panel (b). The pie chart in panel (c) shows the fraction of materials for which the absolute size of the exchange coupling is increased/decreased by the Hubbard-U correction. The pie chart in panel (d) shows how the distribution of materials according to the sign of the exchange coupling predicted by the two methods. All PBE+U calculations were performed with a Hubbard-correction of 4 eV.}
    \label{fig:exchange}
\end{figure*}

Based on these results we conclude that PBE yields more reliable structures than PBE+U (except for the cases where the two methods yields qualitatively different electronic structures, i.e. magnetic/non-megantic or metallic/non-metallic). Therefore we base all subsequent calculations and analysis on the PBE relaxed monolayers.

\subsection{Band gaps}
Figure \ref{fig:gaps} provides a comparative overview of the band gaps calculated with the PBE and PBE+U methods, respectively. The criteria used to select the 638 monolayers included in the comparison are illustrated by the screening funnel in panel (b). Note that only materials with a finite PBE+U band gap are included. Of these 134 have a vanishing PBE band gap. We did not find any materials showing the opposite behavior, i.e. insulator-to-metal transition driven by the U-correction. 

It is clear from the main panel (a) and the inset showing the difference in the band gap obtained with the two methods, that the Hubbard-U correction generally increases the band gap. In fact, the band gap increases in about 90\% of the cases (see panel (c)) and in the remaining 10\% of the cases, the decrease in the band gap is mostly below 0.1 eV.

We note that there is no clear trends in the data set. The change in the band gap varies between 0 and 3.5 eV, although the typical effect of U is to increase the PBE band gap by 0-0.5 eV.

\color{black}Although, as mentioned earlier, the goal of this work is not to assess whether U-corrections improve the physical properties, we find it relevant to compare the calculated and experimental band gaps for a few well known 2D materials. The band gap of CrI$_3$ is increased from 0.89 eV (U=0 eV) to 1.2 eV (U=4.0 eV) bringing it in excellent agreement with experimental reports \cite{huang2017layer}. Adding the U-correction also increases the band gap of CrSBr from 0.45 eV to 0.99 eV, which improves the agreement with the experimental value of 1.8 eV, although the value is still significantly too low\cite{wilson2021interlayer}. For the case of VI$_3$, the Hubbard U opens a small band gap of 0.016 eV in an otherwise metallic structure. However, the magnitude of the calculated gap is considerably smaller than the value of 0.6 eV reported by Hovančík et al\cite{kong2019vi3}. The reason for the small band gap could be related to the fully unquenched orbital angular momentum reported in this compound\cite{hovancik2023large}. This is expected to influence the electronic structure and band gap, but arises as a consequence of spin-orbit coupling, which is not included self-consistently in the present calculations.\color{black}

\subsection{Magnetic moments}
Figure \ref{fig:magmom} shows a comparative overview of the total magnetic moments calculated with the PBE and PBE+U methods, respectively. In total 430 monolayers are included in the comparison and the criteria used to select them are summarised in panel (b). We are excluding materials that exhibits a metal-insulator transition, since these are expected to exhibit a change in magnetic moments as explained earlier. Instead, we focus on materials that are insulating in PBE as well as PBE+U and any change in magnetic moments can be assigned to a change in oxidation state or Hund's rule occupation of the $d$-bands. The magnetic moments are always evaluated for the ferromagnetic phase - also in cases where an anti-ferromagnetic phase has lower energy (see next section). Since we only consider materials with a finite band gap, the total magnetic moments of the unit cell can only take values $N\mu_\mathrm{B}$, where $N$ is an integer. From the main panel and the inset showing the difference in the magnetic moment calculated with the two methods, it is clear that the U-correction has negligible effect on the size of the magnetic moment. In fact there are only nine cases where the U-correction leads to a different magnetic moment (always larger).

\subsection{Magnetic anisotropy}
In Figure \ref{fig:anisotropy} we present an overview of the magnetic anisotropy with and without Hubbard corrections. Again, we only include materials that are non-metallic both with and without U since any metal-to-insulator transition will result in drastic changes that may obscure possible trends. The magnetic anisotropy is evaluated from the energy difference between spins aligned orthogonal to the plane and spins aligned in the plane (along the $x$-axis). These energy differences are evaluated from the magnetic force theorem using non-self-consistent spin-orbit coupling\cite{Olsen2016a}.
A negative anisotropy signifies an easy-axis orthogonal to the plane whereas a positive value implies either an easy-plane or an easy-axis aligned with the atomic plane.

From panel (a) we observe that the magnitude of magnetic anisotropies generally decreases upon inclusion of U. This behaviour is expected due to the increase of orbital localization, which reduces the crystal field effects that drive the magnetic anisotropy. In panel (d) we show the number of materials where the easy-axis changes to or from the out-of-plane direction. In most cases PBE and PBE+U agrees on the sign of the anisotropy (and thus the easy-axis(easy-plane)).

\subsection{Exchange coupling}
In Figure \ref{fig:exchange} we show an overview of the nearest neighbor exchange constants evaluated with and without Hubbard corrections. These were evaluated from energy differences between states of aligned and anti-aligned spins and essentially determines whether the ground state is ferromagnetic ($J>0$) or anti-ferromagnetic ($J<0$) while the magnitude provides a measure of the strength of the magnetic interactions. Again, we only include materials that exhibit a finite band gap in both PBE and PBE+U (panel (b)), since any compounds exhibiting a metal-to-insulator transition are expected to show large differences and thus disturbing the general trends. 

From panels (a) and (c) it is clear that the anti-ferromagnetic exchange interactions decrease upon inclusion of the Hubbard U while the ferromagnetic interactions do not seem to exhibit an obvious trend. The anti-ferromagnetic interactions typically originate from superexchange, which scales as $t^2/U$ and are expected to decrease as a result of orbital localization. On the other hand, ferromagnetic interactions may originate from either superexchange or direct exchange. While the direct exchange is also expected to decrease when orbitals become more localized the dependence of U may be different than for the case of superexchange. Although superexchange is expected to be most prominent in insulators both exchange mechanisms always contribute to the magnetic interactions and direct exchange may be significant in some cases. Thus if the overall magnetic interactions receive contributions from direct exchange (favoring ferromagnetic order) and superexchange (favoring antiferromagnetic order) the total magnetic interaction may increase if the magnitude of superexchange decreases more than the magnitude of direct exchange when Hubbard corrections are included.

In panel (d) we show the number of materials where the nearest neighbor exchange changes sign in the PBE+U description compared to PBE. For the vast majority of materials the sign is the same in PBE and PBE+U. Most of the materials where a sign change is observed change from being anti-ferromagnetic (with PBE) to ferromagnetic (with PBE+U). Again, this trend is expected if we assume that (ferromagnetic) direct exchange is less affected by U than (anti-ferromagnetic) superexchange. However, it should be stressed that the sign of the nearest neighbor exchange does not rigorously determine the magnetic ground state and either negative or positive values may lead to a spiral order, which occur commonly in these materials\cite{sødequist2023magnetic}.

\section{Conclusions}
We have investigated the effect of Hubbard U-corrections in DFT calculations of electronic and magnetic properties of a large set of 2D materials containing $3d$ transition metal elements. Our calculations show that while the band gap can depend strongly on U (most often, but not always increasing with U), the total magnetic moment is essentially independent of U when both PBE and PBE+U yields a finite band gap. For materials undergoing a metal-insulator transition upon inclusion of U, the magnetic properties can change in drastic and unpredictable ways and for that reason such materials were excluded from the comparison. Both the nearest neighbor exchange couplings, in particular anti-ferromagnetic ones ($J<0$), and the magnetic anisotropies are decreased in size by the U-correction. Both effects could be explained by the increased localisation of the $3d$ states carrying the magnetic moments. From this we conclude that the use of Hubbard-U corrections will lead to lower predicted Curie temperatures in 2D materials. All the results of the study are available in the C2DB database.

\section{Acknowledgements}
We acknowledge funding from the European Research Council (ERC) under the European Union’s Horizon 2020 research and innovation program Grant No. 773122 (LIMA) and Grant agreement No. 951786 (NOMAD CoE). K. S. T. is a Villum Investigator supported by VILLUM FONDEN (grant no. 37789).


\begin{thebibliography}{59}%
\makeatletter
\providecommand \@ifxundefined [1]{%
 \@ifx{#1\undefined}
}%
\providecommand \@ifnum [1]{%
 \ifnum #1\expandafter \@firstoftwo
 \else \expandafter \@secondoftwo
 \fi
}%
\providecommand \@ifx [1]{%
 \ifx #1\expandafter \@firstoftwo
 \else \expandafter \@secondoftwo
 \fi
}%
\providecommand \natexlab [1]{#1}%
\providecommand \enquote  [1]{``#1''}%
\providecommand \bibnamefont  [1]{#1}%
\providecommand \bibfnamefont [1]{#1}%
\providecommand \citenamefont [1]{#1}%
\providecommand \href@noop [0]{\@secondoftwo}%
\providecommand \href [0]{\begingroup \@sanitize@url \@href}%
\providecommand \@href[1]{\@@startlink{#1}\@@href}%
\providecommand \@@href[1]{\endgroup#1\@@endlink}%
\providecommand \@sanitize@url [0]{\catcode `\\12\catcode `\$12\catcode `\&12\catcode `\#12\catcode `\^12\catcode `\_12\catcode `\%12\relax}%
\providecommand \@@startlink[1]{}%
\providecommand \@@endlink[0]{}%
\providecommand \url  [0]{\begingroup\@sanitize@url \@url }%
\providecommand \@url [1]{\endgroup\@href {#1}{\urlprefix }}%
\providecommand \urlprefix  [0]{URL }%
\providecommand \Eprint [0]{\href }%
\providecommand \doibase [0]{http://dx.doi.org/}%
\providecommand \selectlanguage [0]{\@gobble}%
\providecommand \bibinfo  [0]{\@secondoftwo}%
\providecommand \bibfield  [0]{\@secondoftwo}%
\providecommand \translation [1]{[#1]}%
\providecommand \BibitemOpen [0]{}%
\providecommand \bibitemStop [0]{}%
\providecommand \bibitemNoStop [0]{.\EOS\space}%
\providecommand \EOS [0]{\spacefactor3000\relax}%
\providecommand \BibitemShut  [1]{\csname bibitem#1\endcsname}%
\let\auto@bib@innerbib\@empty
\bibitem [{\citenamefont {Hubbard}(1964)}]{hubbard1964electron}%
  \BibitemOpen
  \bibfield  {author} {\bibinfo {author} {\bibfnamefont {J.}~\bibnamefont {Hubbard}},\ }\href@noop {} {\bibfield  {journal} {\bibinfo  {journal} {Proceedings of the Royal Society of London. Series A. Mathematical and Physical Sciences}\ }\textbf {\bibinfo {volume} {277}},\ \bibinfo {pages} {237} (\bibinfo {year} {1964})}\BibitemShut {NoStop}%
\bibitem [{\citenamefont {Mancini}(2000)}]{mancini2000mott}%
  \BibitemOpen
  \bibfield  {author} {\bibinfo {author} {\bibfnamefont {F.}~\bibnamefont {Mancini}},\ }\href@noop {} {\bibfield  {journal} {\bibinfo  {journal} {Europhysics Letters}\ }\textbf {\bibinfo {volume} {50}},\ \bibinfo {pages} {229} (\bibinfo {year} {2000})}\BibitemShut {NoStop}%
\bibitem [{\citenamefont {Mott}(1968)}]{mott1968metal}%
  \BibitemOpen
  \bibfield  {author} {\bibinfo {author} {\bibfnamefont {N.~F.}\ \bibnamefont {Mott}},\ }\href@noop {} {\bibfield  {journal} {\bibinfo  {journal} {Reviews of Modern Physics}\ }\textbf {\bibinfo {volume} {40}},\ \bibinfo {pages} {677} (\bibinfo {year} {1968})}\BibitemShut {NoStop}%
\bibitem [{\citenamefont {Stoner}(1938)}]{stoner1938collective}%
  \BibitemOpen
  \bibfield  {author} {\bibinfo {author} {\bibfnamefont {E.~C.}\ \bibnamefont {Stoner}},\ }\href@noop {} {\bibfield  {journal} {\bibinfo  {journal} {Proceedings of the Royal Society of London. Series A. Mathematical and Physical Sciences}\ }\textbf {\bibinfo {volume} {165}},\ \bibinfo {pages} {372} (\bibinfo {year} {1938})}\BibitemShut {NoStop}%
\bibitem [{\citenamefont {Metzner}\ and\ \citenamefont {Vollhardt}(1989)}]{metzner1989correlated}%
  \BibitemOpen
  \bibfield  {author} {\bibinfo {author} {\bibfnamefont {W.}~\bibnamefont {Metzner}}\ and\ \bibinfo {author} {\bibfnamefont {D.}~\bibnamefont {Vollhardt}},\ }\href@noop {} {\bibfield  {journal} {\bibinfo  {journal} {Physical Review Letters}\ }\textbf {\bibinfo {volume} {62}},\ \bibinfo {pages} {324} (\bibinfo {year} {1989})}\BibitemShut {NoStop}%
\bibitem [{\citenamefont {Georges}\ and\ \citenamefont {Kotliar}(1992)}]{georges1992hubbard}%
  \BibitemOpen
  \bibfield  {author} {\bibinfo {author} {\bibfnamefont {A.}~\bibnamefont {Georges}}\ and\ \bibinfo {author} {\bibfnamefont {G.}~\bibnamefont {Kotliar}},\ }\href@noop {} {\bibfield  {journal} {\bibinfo  {journal} {Physical Review B}\ }\textbf {\bibinfo {volume} {45}},\ \bibinfo {pages} {6479} (\bibinfo {year} {1992})}\BibitemShut {NoStop}%
\bibitem [{\citenamefont {Georges}\ \emph {et~al.}(1996)\citenamefont {Georges}, \citenamefont {Kotliar}, \citenamefont {Krauth},\ and\ \citenamefont {Rozenberg}}]{georges1996dynamical}%
  \BibitemOpen
  \bibfield  {author} {\bibinfo {author} {\bibfnamefont {A.}~\bibnamefont {Georges}}, \bibinfo {author} {\bibfnamefont {G.}~\bibnamefont {Kotliar}}, \bibinfo {author} {\bibfnamefont {W.}~\bibnamefont {Krauth}}, \ and\ \bibinfo {author} {\bibfnamefont {M.~J.}\ \bibnamefont {Rozenberg}},\ }\href@noop {} {\bibfield  {journal} {\bibinfo  {journal} {Reviews of Modern Physics}\ }\textbf {\bibinfo {volume} {68}},\ \bibinfo {pages} {13} (\bibinfo {year} {1996})}\BibitemShut {NoStop}%
\bibitem [{\citenamefont {Yang}\ \emph {et~al.}(2000)\citenamefont {Yang}, \citenamefont {Zhang},\ and\ \citenamefont {Ayers}}]{yang2000degenerate}%
  \BibitemOpen
  \bibfield  {author} {\bibinfo {author} {\bibfnamefont {W.}~\bibnamefont {Yang}}, \bibinfo {author} {\bibfnamefont {Y.}~\bibnamefont {Zhang}}, \ and\ \bibinfo {author} {\bibfnamefont {P.~W.}\ \bibnamefont {Ayers}},\ }\href@noop {} {\bibfield  {journal} {\bibinfo  {journal} {Physical Review Letters}\ }\textbf {\bibinfo {volume} {84}},\ \bibinfo {pages} {5172} (\bibinfo {year} {2000})}\BibitemShut {NoStop}%
\bibitem [{\citenamefont {Harl}\ \emph {et~al.}(2010)\citenamefont {Harl}, \citenamefont {Schimka},\ and\ \citenamefont {Kresse}}]{harl2010assessing}%
  \BibitemOpen
  \bibfield  {author} {\bibinfo {author} {\bibfnamefont {J.}~\bibnamefont {Harl}}, \bibinfo {author} {\bibfnamefont {L.}~\bibnamefont {Schimka}}, \ and\ \bibinfo {author} {\bibfnamefont {G.}~\bibnamefont {Kresse}},\ }\href@noop {} {\bibfield  {journal} {\bibinfo  {journal} {Physical Review B}\ }\textbf {\bibinfo {volume} {81}},\ \bibinfo {pages} {115126} (\bibinfo {year} {2010})}\BibitemShut {NoStop}%
\bibitem [{\citenamefont {Olsen}\ and\ \citenamefont {Thygesen}(2013)}]{olsen2013random}%
  \BibitemOpen
  \bibfield  {author} {\bibinfo {author} {\bibfnamefont {T.}~\bibnamefont {Olsen}}\ and\ \bibinfo {author} {\bibfnamefont {K.~S.}\ \bibnamefont {Thygesen}},\ }\href@noop {} {\bibfield  {journal} {\bibinfo  {journal} {Physical Review B}\ }\textbf {\bibinfo {volume} {87}},\ \bibinfo {pages} {075111} (\bibinfo {year} {2013})}\BibitemShut {NoStop}%
\bibitem [{\citenamefont {Jauho}\ \emph {et~al.}(2015)\citenamefont {Jauho}, \citenamefont {Olsen}, \citenamefont {Bligaard},\ and\ \citenamefont {Thygesen}}]{jauho2015improved}%
  \BibitemOpen
  \bibfield  {author} {\bibinfo {author} {\bibfnamefont {T.~S.}\ \bibnamefont {Jauho}}, \bibinfo {author} {\bibfnamefont {T.}~\bibnamefont {Olsen}}, \bibinfo {author} {\bibfnamefont {T.}~\bibnamefont {Bligaard}}, \ and\ \bibinfo {author} {\bibfnamefont {K.~S.}\ \bibnamefont {Thygesen}},\ }\href@noop {} {\bibfield  {journal} {\bibinfo  {journal} {Physical Review B}\ }\textbf {\bibinfo {volume} {92}},\ \bibinfo {pages} {115140} (\bibinfo {year} {2015})}\BibitemShut {NoStop}%
\bibitem [{\citenamefont {Olsen}\ and\ \citenamefont {Thygesen}(2012)}]{olsen2012extending}%
  \BibitemOpen
  \bibfield  {author} {\bibinfo {author} {\bibfnamefont {T.}~\bibnamefont {Olsen}}\ and\ \bibinfo {author} {\bibfnamefont {K.~S.}\ \bibnamefont {Thygesen}},\ }\href@noop {} {\bibfield  {journal} {\bibinfo  {journal} {Physical Review B}\ }\textbf {\bibinfo {volume} {86}},\ \bibinfo {pages} {081103} (\bibinfo {year} {2012})}\BibitemShut {NoStop}%
\bibitem [{\citenamefont {Gr{\"u}neis}\ \emph {et~al.}(2009)\citenamefont {Gr{\"u}neis}, \citenamefont {Marsman}, \citenamefont {Harl}, \citenamefont {Schimka},\ and\ \citenamefont {Kresse}}]{gruneis2009making}%
  \BibitemOpen
  \bibfield  {author} {\bibinfo {author} {\bibfnamefont {A.}~\bibnamefont {Gr{\"u}neis}}, \bibinfo {author} {\bibfnamefont {M.}~\bibnamefont {Marsman}}, \bibinfo {author} {\bibfnamefont {J.}~\bibnamefont {Harl}}, \bibinfo {author} {\bibfnamefont {L.}~\bibnamefont {Schimka}}, \ and\ \bibinfo {author} {\bibfnamefont {G.}~\bibnamefont {Kresse}},\ }\href@noop {} {\bibfield  {journal} {\bibinfo  {journal} {The Journal of chemical physics}\ }\textbf {\bibinfo {volume} {131}} (\bibinfo {year} {2009})}\BibitemShut {NoStop}%
\bibitem [{\citenamefont {Olsen}\ and\ \citenamefont {Thygesen}(2014)}]{olsen2014accurate}%
  \BibitemOpen
  \bibfield  {author} {\bibinfo {author} {\bibfnamefont {T.}~\bibnamefont {Olsen}}\ and\ \bibinfo {author} {\bibfnamefont {K.~S.}\ \bibnamefont {Thygesen}},\ }\href@noop {} {\bibfield  {journal} {\bibinfo  {journal} {Physical Review Letters}\ }\textbf {\bibinfo {volume} {112}},\ \bibinfo {pages} {203001} (\bibinfo {year} {2014})}\BibitemShut {NoStop}%
\bibitem [{\citenamefont {Olsen}\ \emph {et~al.}(2019)\citenamefont {Olsen}, \citenamefont {Patrick}, \citenamefont {Bates}, \citenamefont {Ruzsinszky},\ and\ \citenamefont {Thygesen}}]{olsen2019beyond}%
  \BibitemOpen
  \bibfield  {author} {\bibinfo {author} {\bibfnamefont {T.}~\bibnamefont {Olsen}}, \bibinfo {author} {\bibfnamefont {C.~E.}\ \bibnamefont {Patrick}}, \bibinfo {author} {\bibfnamefont {J.~E.}\ \bibnamefont {Bates}}, \bibinfo {author} {\bibfnamefont {A.}~\bibnamefont {Ruzsinszky}}, \ and\ \bibinfo {author} {\bibfnamefont {K.~S.}\ \bibnamefont {Thygesen}},\ }\href@noop {} {\bibfield  {journal} {\bibinfo  {journal} {npj Computational Materials}\ }\textbf {\bibinfo {volume} {5}},\ \bibinfo {pages} {106} (\bibinfo {year} {2019})}\BibitemShut {NoStop}%
\bibitem [{\citenamefont {Patrick}\ and\ \citenamefont {Thygesen}(2015)}]{patrick2015adiabatic}%
  \BibitemOpen
  \bibfield  {author} {\bibinfo {author} {\bibfnamefont {C.~E.}\ \bibnamefont {Patrick}}\ and\ \bibinfo {author} {\bibfnamefont {K.~S.}\ \bibnamefont {Thygesen}},\ }\href@noop {} {\bibfield  {journal} {\bibinfo  {journal} {The Journal of Chemical Physics}\ }\textbf {\bibinfo {volume} {143}} (\bibinfo {year} {2015})}\BibitemShut {NoStop}%
\bibitem [{\citenamefont {Sch{\"a}fer}\ \emph {et~al.}(2016)\citenamefont {Sch{\"a}fer}, \citenamefont {Ciuchi}, \citenamefont {Wallerberger}, \citenamefont {Thunstr{\"o}m}, \citenamefont {Gunnarsson}, \citenamefont {Sangiovanni}, \citenamefont {Rohringer},\ and\ \citenamefont {Toschi}}]{schafer2016nonperturbative}%
  \BibitemOpen
  \bibfield  {author} {\bibinfo {author} {\bibfnamefont {T.}~\bibnamefont {Sch{\"a}fer}}, \bibinfo {author} {\bibfnamefont {S.}~\bibnamefont {Ciuchi}}, \bibinfo {author} {\bibfnamefont {M.}~\bibnamefont {Wallerberger}}, \bibinfo {author} {\bibfnamefont {P.}~\bibnamefont {Thunstr{\"o}m}}, \bibinfo {author} {\bibfnamefont {O.}~\bibnamefont {Gunnarsson}}, \bibinfo {author} {\bibfnamefont {G.}~\bibnamefont {Sangiovanni}}, \bibinfo {author} {\bibfnamefont {G.}~\bibnamefont {Rohringer}}, \ and\ \bibinfo {author} {\bibfnamefont {A.}~\bibnamefont {Toschi}},\ }\href@noop {} {\bibfield  {journal} {\bibinfo  {journal} {Physical Review B}\ }\textbf {\bibinfo {volume} {94}},\ \bibinfo {pages} {235108} (\bibinfo {year} {2016})}\BibitemShut {NoStop}%
\bibitem [{\citenamefont {Hohenberg}\ and\ \citenamefont {Kohn}(1964)}]{hohenberg1964inhomogeneous}%
  \BibitemOpen
  \bibfield  {author} {\bibinfo {author} {\bibfnamefont {P.}~\bibnamefont {Hohenberg}}\ and\ \bibinfo {author} {\bibfnamefont {W.}~\bibnamefont {Kohn}},\ }\href@noop {} {\bibfield  {journal} {\bibinfo  {journal} {Physical Review}\ }\textbf {\bibinfo {volume} {136}},\ \bibinfo {pages} {B864} (\bibinfo {year} {1964})}\BibitemShut {NoStop}%
\bibitem [{\citenamefont {Kohn}\ and\ \citenamefont {Sham}(1965)}]{kohn1965self}%
  \BibitemOpen
  \bibfield  {author} {\bibinfo {author} {\bibfnamefont {W.}~\bibnamefont {Kohn}}\ and\ \bibinfo {author} {\bibfnamefont {L.~J.}\ \bibnamefont {Sham}},\ }\href@noop {} {\bibfield  {journal} {\bibinfo  {journal} {Physical review}\ }\textbf {\bibinfo {volume} {140}},\ \bibinfo {pages} {A1133} (\bibinfo {year} {1965})}\BibitemShut {NoStop}%
\bibitem [{\citenamefont {Cohen}\ \emph {et~al.}(2008)\citenamefont {Cohen}, \citenamefont {Mori-S{\'a}nchez},\ and\ \citenamefont {Yang}}]{cohen2008insights}%
  \BibitemOpen
  \bibfield  {author} {\bibinfo {author} {\bibfnamefont {A.~J.}\ \bibnamefont {Cohen}}, \bibinfo {author} {\bibfnamefont {P.}~\bibnamefont {Mori-S{\'a}nchez}}, \ and\ \bibinfo {author} {\bibfnamefont {W.}~\bibnamefont {Yang}},\ }\href@noop {} {\bibfield  {journal} {\bibinfo  {journal} {Science}\ }\textbf {\bibinfo {volume} {321}},\ \bibinfo {pages} {792} (\bibinfo {year} {2008})}\BibitemShut {NoStop}%
\bibitem [{\citenamefont {Perdew}\ and\ \citenamefont {Zunger}(1981)}]{perdew1981self}%
  \BibitemOpen
  \bibfield  {author} {\bibinfo {author} {\bibfnamefont {J.~P.}\ \bibnamefont {Perdew}}\ and\ \bibinfo {author} {\bibfnamefont {A.}~\bibnamefont {Zunger}},\ }\href@noop {} {\bibfield  {journal} {\bibinfo  {journal} {Physical Review B}\ }\textbf {\bibinfo {volume} {23}},\ \bibinfo {pages} {5048} (\bibinfo {year} {1981})}\BibitemShut {NoStop}%
\bibitem [{\citenamefont {Anisimov}\ and\ \citenamefont {Gunnarsson}(1991)}]{anisimov1991density}%
  \BibitemOpen
  \bibfield  {author} {\bibinfo {author} {\bibfnamefont {V.}~\bibnamefont {Anisimov}}\ and\ \bibinfo {author} {\bibfnamefont {O.}~\bibnamefont {Gunnarsson}},\ }\href@noop {} {\bibfield  {journal} {\bibinfo  {journal} {Physical Review B}\ }\textbf {\bibinfo {volume} {43}},\ \bibinfo {pages} {7570} (\bibinfo {year} {1991})}\BibitemShut {NoStop}%
\bibitem [{\citenamefont {Anisimov}\ \emph {et~al.}(1991)\citenamefont {Anisimov}, \citenamefont {Zaanen},\ and\ \citenamefont {Andersen}}]{anisimov1991band}%
  \BibitemOpen
  \bibfield  {author} {\bibinfo {author} {\bibfnamefont {V.~I.}\ \bibnamefont {Anisimov}}, \bibinfo {author} {\bibfnamefont {J.}~\bibnamefont {Zaanen}}, \ and\ \bibinfo {author} {\bibfnamefont {O.~K.}\ \bibnamefont {Andersen}},\ }\href@noop {} {\bibfield  {journal} {\bibinfo  {journal} {Physical Review B}\ }\textbf {\bibinfo {volume} {44}},\ \bibinfo {pages} {943} (\bibinfo {year} {1991})}\BibitemShut {NoStop}%
\bibitem [{\citenamefont {Anisimov}\ \emph {et~al.}(1993)\citenamefont {Anisimov}, \citenamefont {Solovyev}, \citenamefont {Korotin}, \citenamefont {Czy{\.z}yk},\ and\ \citenamefont {Sawatzky}}]{anisimov1993density}%
  \BibitemOpen
  \bibfield  {author} {\bibinfo {author} {\bibfnamefont {V.~I.}\ \bibnamefont {Anisimov}}, \bibinfo {author} {\bibfnamefont {I.}~\bibnamefont {Solovyev}}, \bibinfo {author} {\bibfnamefont {M.}~\bibnamefont {Korotin}}, \bibinfo {author} {\bibfnamefont {M.}~\bibnamefont {Czy{\.z}yk}}, \ and\ \bibinfo {author} {\bibfnamefont {G.}~\bibnamefont {Sawatzky}},\ }\href@noop {} {\bibfield  {journal} {\bibinfo  {journal} {Physical Review B}\ }\textbf {\bibinfo {volume} {48}},\ \bibinfo {pages} {16929} (\bibinfo {year} {1993})}\BibitemShut {NoStop}%
\bibitem [{\citenamefont {Solovyev}\ \emph {et~al.}(1994)\citenamefont {Solovyev}, \citenamefont {Dederichs},\ and\ \citenamefont {Anisimov}}]{solovyev1994corrected}%
  \BibitemOpen
  \bibfield  {author} {\bibinfo {author} {\bibfnamefont {I.}~\bibnamefont {Solovyev}}, \bibinfo {author} {\bibfnamefont {P.}~\bibnamefont {Dederichs}}, \ and\ \bibinfo {author} {\bibfnamefont {V.}~\bibnamefont {Anisimov}},\ }\href@noop {} {\bibfield  {journal} {\bibinfo  {journal} {Physical Review B}\ }\textbf {\bibinfo {volume} {50}},\ \bibinfo {pages} {16861} (\bibinfo {year} {1994})}\BibitemShut {NoStop}%
\bibitem [{\citenamefont {Anisimov}\ \emph {et~al.}(1997)\citenamefont {Anisimov}, \citenamefont {Aryasetiawan},\ and\ \citenamefont {Lichtenstein}}]{anisimov1997first}%
  \BibitemOpen
  \bibfield  {author} {\bibinfo {author} {\bibfnamefont {V.~I.}\ \bibnamefont {Anisimov}}, \bibinfo {author} {\bibfnamefont {F.}~\bibnamefont {Aryasetiawan}}, \ and\ \bibinfo {author} {\bibfnamefont {A.}~\bibnamefont {Lichtenstein}},\ }\href@noop {} {\bibfield  {journal} {\bibinfo  {journal} {Journal of Physics: Condensed Matter}\ }\textbf {\bibinfo {volume} {9}},\ \bibinfo {pages} {767} (\bibinfo {year} {1997})}\BibitemShut {NoStop}%
\bibitem [{\citenamefont {Liechtenstein}\ \emph {et~al.}(1995)\citenamefont {Liechtenstein}, \citenamefont {Anisimov},\ and\ \citenamefont {Zaanen}}]{liechtenstein1995density}%
  \BibitemOpen
  \bibfield  {author} {\bibinfo {author} {\bibfnamefont {A.}~\bibnamefont {Liechtenstein}}, \bibinfo {author} {\bibfnamefont {V.~I.}\ \bibnamefont {Anisimov}}, \ and\ \bibinfo {author} {\bibfnamefont {J.}~\bibnamefont {Zaanen}},\ }\href@noop {} {\bibfield  {journal} {\bibinfo  {journal} {Physical Review B}\ }\textbf {\bibinfo {volume} {52}},\ \bibinfo {pages} {R5467} (\bibinfo {year} {1995})}\BibitemShut {NoStop}%
\bibitem [{\citenamefont {Wang}\ \emph {et~al.}(2006)\citenamefont {Wang}, \citenamefont {Maxisch},\ and\ \citenamefont {Ceder}}]{PhysRevB.73.195107}%
  \BibitemOpen
  \bibfield  {author} {\bibinfo {author} {\bibfnamefont {L.}~\bibnamefont {Wang}}, \bibinfo {author} {\bibfnamefont {T.}~\bibnamefont {Maxisch}}, \ and\ \bibinfo {author} {\bibfnamefont {G.}~\bibnamefont {Ceder}},\ }\href {\doibase 10.1103/PhysRevB.73.195107} {\bibfield  {journal} {\bibinfo  {journal} {Phys. Rev. B}\ }\textbf {\bibinfo {volume} {73}},\ \bibinfo {pages} {195107} (\bibinfo {year} {2006})}\BibitemShut {NoStop}%
\bibitem [{\citenamefont {Cococcioni}\ and\ \citenamefont {de~Gironcoli}(2005)}]{PhysRevB.71.035105}%
  \BibitemOpen
  \bibfield  {author} {\bibinfo {author} {\bibfnamefont {M.}~\bibnamefont {Cococcioni}}\ and\ \bibinfo {author} {\bibfnamefont {S.}~\bibnamefont {de~Gironcoli}},\ }\href {\doibase 10.1103/PhysRevB.71.035105} {\bibfield  {journal} {\bibinfo  {journal} {Phys. Rev. B}\ }\textbf {\bibinfo {volume} {71}},\ \bibinfo {pages} {035105} (\bibinfo {year} {2005})}\BibitemShut {NoStop}%
\bibitem [{\citenamefont {Yekta}\ \emph {et~al.}(2021)\citenamefont {Yekta}, \citenamefont {Hadipour}, \citenamefont {\ifmmode \mbox{\c{S}}\else \c{S}\fi{}a\ifmmode \mbox{\c{s}}\else \c{s}\fi{}\ifmmode \imath \else \i \fi{}o\ifmmode~\breve{g}\else \u{g}\fi{}lu}, \citenamefont {Friedrich}, \citenamefont {Jafari}, \citenamefont {Bl\"ugel},\ and\ \citenamefont {Mertig}}]{PhysRevMaterials.5.034001}%
  \BibitemOpen
  \bibfield  {author} {\bibinfo {author} {\bibfnamefont {Y.}~\bibnamefont {Yekta}}, \bibinfo {author} {\bibfnamefont {H.}~\bibnamefont {Hadipour}}, \bibinfo {author} {\bibfnamefont {E.}~\bibnamefont {\ifmmode \mbox{\c{S}}\else \c{S}\fi{}a\ifmmode \mbox{\c{s}}\else \c{s}\fi{}\ifmmode \imath \else \i \fi{}o\ifmmode~\breve{g}\else \u{g}\fi{}lu}}, \bibinfo {author} {\bibfnamefont {C.}~\bibnamefont {Friedrich}}, \bibinfo {author} {\bibfnamefont {S.~A.}\ \bibnamefont {Jafari}}, \bibinfo {author} {\bibfnamefont {S.}~\bibnamefont {Bl\"ugel}}, \ and\ \bibinfo {author} {\bibfnamefont {I.}~\bibnamefont {Mertig}},\ }\href {\doibase 10.1103/PhysRevMaterials.5.034001} {\bibfield  {journal} {\bibinfo  {journal} {Phys. Rev. Mater.}\ }\textbf {\bibinfo {volume} {5}},\ \bibinfo {pages} {034001} (\bibinfo {year} {2021})}\BibitemShut {NoStop}%
\bibitem [{\citenamefont {Himmetoglu}\ \emph {et~al.}(2014)\citenamefont {Himmetoglu}, \citenamefont {Floris}, \citenamefont {De~Gironcoli},\ and\ \citenamefont {Cococcioni}}]{himmetoglu2014hubbard}%
  \BibitemOpen
  \bibfield  {author} {\bibinfo {author} {\bibfnamefont {B.}~\bibnamefont {Himmetoglu}}, \bibinfo {author} {\bibfnamefont {A.}~\bibnamefont {Floris}}, \bibinfo {author} {\bibfnamefont {S.}~\bibnamefont {De~Gironcoli}}, \ and\ \bibinfo {author} {\bibfnamefont {M.}~\bibnamefont {Cococcioni}},\ }\href@noop {} {\bibfield  {journal} {\bibinfo  {journal} {International Journal of Quantum Chemistry}\ }\textbf {\bibinfo {volume} {114}},\ \bibinfo {pages} {14} (\bibinfo {year} {2014})}\BibitemShut {NoStop}%
\bibitem [{\citenamefont {Vaugier}\ \emph {et~al.}(2012)\citenamefont {Vaugier}, \citenamefont {Jiang},\ and\ \citenamefont {Biermann}}]{vaugier2012hubbard}%
  \BibitemOpen
  \bibfield  {author} {\bibinfo {author} {\bibfnamefont {L.}~\bibnamefont {Vaugier}}, \bibinfo {author} {\bibfnamefont {H.}~\bibnamefont {Jiang}}, \ and\ \bibinfo {author} {\bibfnamefont {S.}~\bibnamefont {Biermann}},\ }\href@noop {} {\bibfield  {journal} {\bibinfo  {journal} {Physical Review B}\ }\textbf {\bibinfo {volume} {86}},\ \bibinfo {pages} {165105} (\bibinfo {year} {2012})}\BibitemShut {NoStop}%
\bibitem [{\citenamefont {Orhan}\ and\ \citenamefont {O'Regan}(2020)}]{orhan2020first}%
  \BibitemOpen
  \bibfield  {author} {\bibinfo {author} {\bibfnamefont {O.~K.}\ \bibnamefont {Orhan}}\ and\ \bibinfo {author} {\bibfnamefont {D.~D.}\ \bibnamefont {O'Regan}},\ }\href@noop {} {\bibfield  {journal} {\bibinfo  {journal} {Physical Review B}\ }\textbf {\bibinfo {volume} {101}},\ \bibinfo {pages} {245137} (\bibinfo {year} {2020})}\BibitemShut {NoStop}%
\bibitem [{\citenamefont {Timrov}\ \emph {et~al.}(2018)\citenamefont {Timrov}, \citenamefont {Marzari},\ and\ \citenamefont {Cococcioni}}]{timrov2018hubbard}%
  \BibitemOpen
  \bibfield  {author} {\bibinfo {author} {\bibfnamefont {I.}~\bibnamefont {Timrov}}, \bibinfo {author} {\bibfnamefont {N.}~\bibnamefont {Marzari}}, \ and\ \bibinfo {author} {\bibfnamefont {M.}~\bibnamefont {Cococcioni}},\ }\href@noop {} {\bibfield  {journal} {\bibinfo  {journal} {Physical Review B}\ }\textbf {\bibinfo {volume} {98}},\ \bibinfo {pages} {085127} (\bibinfo {year} {2018})}\BibitemShut {NoStop}%
\bibitem [{\citenamefont {Shenton}\ \emph {et~al.}(2017)\citenamefont {Shenton}, \citenamefont {Bowler},\ and\ \citenamefont {Cheah}}]{shenton2017effects}%
  \BibitemOpen
  \bibfield  {author} {\bibinfo {author} {\bibfnamefont {J.~K.}\ \bibnamefont {Shenton}}, \bibinfo {author} {\bibfnamefont {D.~R.}\ \bibnamefont {Bowler}}, \ and\ \bibinfo {author} {\bibfnamefont {W.~L.}\ \bibnamefont {Cheah}},\ }\href@noop {} {\bibfield  {journal} {\bibinfo  {journal} {Journal of Physics: Condensed Matter}\ }\textbf {\bibinfo {volume} {29}},\ \bibinfo {pages} {445501} (\bibinfo {year} {2017})}\BibitemShut {NoStop}%
\bibitem [{\citenamefont {Tancogne-Dejean}\ \emph {et~al.}(2018)\citenamefont {Tancogne-Dejean}, \citenamefont {Sentef},\ and\ \citenamefont {Rubio}}]{tancogne2018ultrafast}%
  \BibitemOpen
  \bibfield  {author} {\bibinfo {author} {\bibfnamefont {N.}~\bibnamefont {Tancogne-Dejean}}, \bibinfo {author} {\bibfnamefont {M.~A.}\ \bibnamefont {Sentef}}, \ and\ \bibinfo {author} {\bibfnamefont {A.}~\bibnamefont {Rubio}},\ }\href@noop {} {\bibfield  {journal} {\bibinfo  {journal} {Physical review letters}\ }\textbf {\bibinfo {volume} {121}},\ \bibinfo {pages} {097402} (\bibinfo {year} {2018})}\BibitemShut {NoStop}%
\bibitem [{\citenamefont {Agapito}\ \emph {et~al.}(2015)\citenamefont {Agapito}, \citenamefont {Curtarolo},\ and\ \citenamefont {Buongiorno~Nardelli}}]{agapito2015reformulation}%
  \BibitemOpen
  \bibfield  {author} {\bibinfo {author} {\bibfnamefont {L.~A.}\ \bibnamefont {Agapito}}, \bibinfo {author} {\bibfnamefont {S.}~\bibnamefont {Curtarolo}}, \ and\ \bibinfo {author} {\bibfnamefont {M.}~\bibnamefont {Buongiorno~Nardelli}},\ }\href@noop {} {\bibfield  {journal} {\bibinfo  {journal} {Physical Review X}\ }\textbf {\bibinfo {volume} {5}},\ \bibinfo {pages} {011006} (\bibinfo {year} {2015})}\BibitemShut {NoStop}%
\bibitem [{\citenamefont {Huang}\ \emph {et~al.}(2020)\citenamefont {Huang}, \citenamefont {Lee}, \citenamefont {Son}, \citenamefont {Supka},\ and\ \citenamefont {Liu}}]{huang2020first}%
  \BibitemOpen
  \bibfield  {author} {\bibinfo {author} {\bibfnamefont {J.}~\bibnamefont {Huang}}, \bibinfo {author} {\bibfnamefont {S.-H.}\ \bibnamefont {Lee}}, \bibinfo {author} {\bibfnamefont {Y.-W.}\ \bibnamefont {Son}}, \bibinfo {author} {\bibfnamefont {A.}~\bibnamefont {Supka}}, \ and\ \bibinfo {author} {\bibfnamefont {S.}~\bibnamefont {Liu}},\ }\href@noop {} {\bibfield  {journal} {\bibinfo  {journal} {Physical Review B}\ }\textbf {\bibinfo {volume} {102}},\ \bibinfo {pages} {165157} (\bibinfo {year} {2020})}\BibitemShut {NoStop}%
\bibitem [{\citenamefont {Pasquier}\ and\ \citenamefont {Yazyev}(2022)}]{pasquier2022ab}%
  \BibitemOpen
  \bibfield  {author} {\bibinfo {author} {\bibfnamefont {D.}~\bibnamefont {Pasquier}}\ and\ \bibinfo {author} {\bibfnamefont {O.~V.}\ \bibnamefont {Yazyev}},\ }\href@noop {} {\bibfield  {journal} {\bibinfo  {journal} {Physical Review B}\ }\textbf {\bibinfo {volume} {105}},\ \bibinfo {pages} {L081106} (\bibinfo {year} {2022})}\BibitemShut {NoStop}%
\bibitem [{\citenamefont {Das}\ \emph {et~al.}(2019)\citenamefont {Das}, \citenamefont {Di~Liberto}, \citenamefont {Tosoni},\ and\ \citenamefont {Pacchioni}}]{das2019band}%
  \BibitemOpen
  \bibfield  {author} {\bibinfo {author} {\bibfnamefont {T.}~\bibnamefont {Das}}, \bibinfo {author} {\bibfnamefont {G.}~\bibnamefont {Di~Liberto}}, \bibinfo {author} {\bibfnamefont {S.}~\bibnamefont {Tosoni}}, \ and\ \bibinfo {author} {\bibfnamefont {G.}~\bibnamefont {Pacchioni}},\ }\href@noop {} {\bibfield  {journal} {\bibinfo  {journal} {Journal of chemical theory and computation}\ }\textbf {\bibinfo {volume} {15}},\ \bibinfo {pages} {6294} (\bibinfo {year} {2019})}\BibitemShut {NoStop}%
\bibitem [{\citenamefont {Li}\ \emph {et~al.}(2020)\citenamefont {Li}, \citenamefont {Zhang},\ and\ \citenamefont {Zhang}}]{li2020high}%
  \BibitemOpen
  \bibfield  {author} {\bibinfo {author} {\bibfnamefont {X.}~\bibnamefont {Li}}, \bibinfo {author} {\bibfnamefont {Z.}~\bibnamefont {Zhang}}, \ and\ \bibinfo {author} {\bibfnamefont {H.}~\bibnamefont {Zhang}},\ }\href@noop {} {\bibfield  {journal} {\bibinfo  {journal} {Nanoscale Advances}\ }\textbf {\bibinfo {volume} {2}},\ \bibinfo {pages} {495} (\bibinfo {year} {2020})}\BibitemShut {NoStop}%
\bibitem [{\citenamefont {Cudazzo}\ \emph {et~al.}(2011)\citenamefont {Cudazzo}, \citenamefont {Tokatly},\ and\ \citenamefont {Rubio}}]{cudazzo2011dielectric}%
  \BibitemOpen
  \bibfield  {author} {\bibinfo {author} {\bibfnamefont {P.}~\bibnamefont {Cudazzo}}, \bibinfo {author} {\bibfnamefont {I.~V.}\ \bibnamefont {Tokatly}}, \ and\ \bibinfo {author} {\bibfnamefont {A.}~\bibnamefont {Rubio}},\ }\href@noop {} {\bibfield  {journal} {\bibinfo  {journal} {Physical Review B}\ }\textbf {\bibinfo {volume} {84}},\ \bibinfo {pages} {085406} (\bibinfo {year} {2011})}\BibitemShut {NoStop}%
\bibitem [{\citenamefont {H{\"u}ser}\ \emph {et~al.}(2013)\citenamefont {H{\"u}ser}, \citenamefont {Olsen},\ and\ \citenamefont {Thygesen}}]{huser2013dielectric}%
  \BibitemOpen
  \bibfield  {author} {\bibinfo {author} {\bibfnamefont {F.}~\bibnamefont {H{\"u}ser}}, \bibinfo {author} {\bibfnamefont {T.}~\bibnamefont {Olsen}}, \ and\ \bibinfo {author} {\bibfnamefont {K.~S.}\ \bibnamefont {Thygesen}},\ }\href@noop {} {\bibfield  {journal} {\bibinfo  {journal} {Physical Review B}\ }\textbf {\bibinfo {volume} {88}},\ \bibinfo {pages} {245309} (\bibinfo {year} {2013})}\BibitemShut {NoStop}%
\bibitem [{\citenamefont {Saal}\ \emph {et~al.}(2013)\citenamefont {Saal}, \citenamefont {Kirklin}, \citenamefont {Aykol}, \citenamefont {Meredig},\ and\ \citenamefont {Wolverton}}]{Saal2013}%
  \BibitemOpen
  \bibfield  {author} {\bibinfo {author} {\bibfnamefont {J.~E.}\ \bibnamefont {Saal}}, \bibinfo {author} {\bibfnamefont {S.}~\bibnamefont {Kirklin}}, \bibinfo {author} {\bibfnamefont {M.}~\bibnamefont {Aykol}}, \bibinfo {author} {\bibfnamefont {B.}~\bibnamefont {Meredig}}, \ and\ \bibinfo {author} {\bibfnamefont {C.}~\bibnamefont {Wolverton}},\ }\href {\doibase 10.1007/s11837-013-0755-4} {\bibfield  {journal} {\bibinfo  {journal} {JOM}\ }\textbf {\bibinfo {volume} {65}},\ \bibinfo {pages} {1501} (\bibinfo {year} {2013})}\BibitemShut {NoStop}%
\bibitem [{\citenamefont {Haastrup}\ \emph {et~al.}(2018)\citenamefont {Haastrup}, \citenamefont {Strange}, \citenamefont {Pandey}, \citenamefont {Deilmann}, \citenamefont {Schmidt}, \citenamefont {Hinsche}, \citenamefont {Gjerding}, \citenamefont {Torelli}, \citenamefont {Larsen}, \citenamefont {Riis-Jensen} \emph {et~al.}}]{haastrup2018computational}%
  \BibitemOpen
  \bibfield  {author} {\bibinfo {author} {\bibfnamefont {S.}~\bibnamefont {Haastrup}}, \bibinfo {author} {\bibfnamefont {M.}~\bibnamefont {Strange}}, \bibinfo {author} {\bibfnamefont {M.}~\bibnamefont {Pandey}}, \bibinfo {author} {\bibfnamefont {T.}~\bibnamefont {Deilmann}}, \bibinfo {author} {\bibfnamefont {P.~S.}\ \bibnamefont {Schmidt}}, \bibinfo {author} {\bibfnamefont {N.~F.}\ \bibnamefont {Hinsche}}, \bibinfo {author} {\bibfnamefont {M.~N.}\ \bibnamefont {Gjerding}}, \bibinfo {author} {\bibfnamefont {D.}~\bibnamefont {Torelli}}, \bibinfo {author} {\bibfnamefont {P.~M.}\ \bibnamefont {Larsen}}, \bibinfo {author} {\bibfnamefont {A.~C.}\ \bibnamefont {Riis-Jensen}},  \emph {et~al.},\ }\href@noop {} {\bibfield  {journal} {\bibinfo  {journal} {2D Materials}\ }\textbf {\bibinfo {volume} {5}},\ \bibinfo {pages} {042002} (\bibinfo {year} {2018})}\BibitemShut {NoStop}%
\bibitem [{\citenamefont {Gjerding}\ \emph {et~al.}(2021{\natexlab{a}})\citenamefont {Gjerding}, \citenamefont {Taghizadeh}, \citenamefont {Rasmussen}, \citenamefont {Ali}, \citenamefont {Bertoldo}, \citenamefont {Deilmann}, \citenamefont {Kn{\o}sgaard}, \citenamefont {Kruse}, \citenamefont {Larsen}, \citenamefont {Manti} \emph {et~al.}}]{gjerding2021recent}%
  \BibitemOpen
  \bibfield  {author} {\bibinfo {author} {\bibfnamefont {M.~N.}\ \bibnamefont {Gjerding}}, \bibinfo {author} {\bibfnamefont {A.}~\bibnamefont {Taghizadeh}}, \bibinfo {author} {\bibfnamefont {A.}~\bibnamefont {Rasmussen}}, \bibinfo {author} {\bibfnamefont {S.}~\bibnamefont {Ali}}, \bibinfo {author} {\bibfnamefont {F.}~\bibnamefont {Bertoldo}}, \bibinfo {author} {\bibfnamefont {T.}~\bibnamefont {Deilmann}}, \bibinfo {author} {\bibfnamefont {N.~R.}\ \bibnamefont {Kn{\o}sgaard}}, \bibinfo {author} {\bibfnamefont {M.}~\bibnamefont {Kruse}}, \bibinfo {author} {\bibfnamefont {A.~H.}\ \bibnamefont {Larsen}}, \bibinfo {author} {\bibfnamefont {S.}~\bibnamefont {Manti}},  \emph {et~al.},\ }\href@noop {} {\bibfield  {journal} {\bibinfo  {journal} {2D Materials}\ }\textbf {\bibinfo {volume} {8}},\ \bibinfo {pages} {044002} (\bibinfo {year} {2021}{\natexlab{a}})}\BibitemShut {NoStop}%
\bibitem [{\citenamefont {Gjerding}\ \emph {et~al.}(2021{\natexlab{b}})\citenamefont {Gjerding}, \citenamefont {Skovhus}, \citenamefont {Rasmussen}, \citenamefont {Bertoldo}, \citenamefont {Larsen}, \citenamefont {Mortensen},\ and\ \citenamefont {Thygesen}}]{gjerding2021atomic}%
  \BibitemOpen
  \bibfield  {author} {\bibinfo {author} {\bibfnamefont {M.}~\bibnamefont {Gjerding}}, \bibinfo {author} {\bibfnamefont {T.}~\bibnamefont {Skovhus}}, \bibinfo {author} {\bibfnamefont {A.}~\bibnamefont {Rasmussen}}, \bibinfo {author} {\bibfnamefont {F.}~\bibnamefont {Bertoldo}}, \bibinfo {author} {\bibfnamefont {A.~H.}\ \bibnamefont {Larsen}}, \bibinfo {author} {\bibfnamefont {J.~J.}\ \bibnamefont {Mortensen}}, \ and\ \bibinfo {author} {\bibfnamefont {K.~S.}\ \bibnamefont {Thygesen}},\ }\href@noop {} {\bibfield  {journal} {\bibinfo  {journal} {Computational Materials Science}\ }\textbf {\bibinfo {volume} {199}},\ \bibinfo {pages} {110731} (\bibinfo {year} {2021}{\natexlab{b}})}\BibitemShut {NoStop}%
\bibitem [{\citenamefont {Mortensen}\ \emph {et~al.}(2020)\citenamefont {Mortensen}, \citenamefont {Gjerding},\ and\ \citenamefont {Thygesen}}]{mortensen2020myqueue}%
  \BibitemOpen
  \bibfield  {author} {\bibinfo {author} {\bibfnamefont {J.~J.}\ \bibnamefont {Mortensen}}, \bibinfo {author} {\bibfnamefont {M.}~\bibnamefont {Gjerding}}, \ and\ \bibinfo {author} {\bibfnamefont {K.~S.}\ \bibnamefont {Thygesen}},\ }\href@noop {} {\bibfield  {journal} {\bibinfo  {journal} {Journal of Open Source Software}\ }\textbf {\bibinfo {volume} {5}},\ \bibinfo {pages} {1844} (\bibinfo {year} {2020})}\BibitemShut {NoStop}%
\bibitem [{\citenamefont {Mortensen}\ \emph {et~al.}(2024)\citenamefont {Mortensen}, \citenamefont {Larsen}, \citenamefont {Kuisma}, \citenamefont {Ivanov}, \citenamefont {Taghizadeh}, \citenamefont {Peterson}, \citenamefont {Haldar}, \citenamefont {Dohn}, \citenamefont {Sch{\"a}fer}, \citenamefont {J{\'o}nsson} \emph {et~al.}}]{mortensen2024gpaw}%
  \BibitemOpen
  \bibfield  {author} {\bibinfo {author} {\bibfnamefont {J.~J.}\ \bibnamefont {Mortensen}}, \bibinfo {author} {\bibfnamefont {A.~H.}\ \bibnamefont {Larsen}}, \bibinfo {author} {\bibfnamefont {M.}~\bibnamefont {Kuisma}}, \bibinfo {author} {\bibfnamefont {A.~V.}\ \bibnamefont {Ivanov}}, \bibinfo {author} {\bibfnamefont {A.}~\bibnamefont {Taghizadeh}}, \bibinfo {author} {\bibfnamefont {A.}~\bibnamefont {Peterson}}, \bibinfo {author} {\bibfnamefont {A.}~\bibnamefont {Haldar}}, \bibinfo {author} {\bibfnamefont {A.~O.}\ \bibnamefont {Dohn}}, \bibinfo {author} {\bibfnamefont {C.}~\bibnamefont {Sch{\"a}fer}}, \bibinfo {author} {\bibfnamefont {E.~{\"O}.}\ \bibnamefont {J{\'o}nsson}},  \emph {et~al.},\ }\href@noop {} {\bibfield  {journal} {\bibinfo  {journal} {The Journal of Chemical Physics}\ }\textbf {\bibinfo {volume} {160}} (\bibinfo {year} {2024})}\BibitemShut {NoStop}%
\bibitem [{\citenamefont {Enkovaara}\ \emph {et~al.}(2010)\citenamefont {Enkovaara}, \citenamefont {Rostgaard}, \citenamefont {Mortensen}, \citenamefont {Chen}, \citenamefont {Du{\l}ak}, \citenamefont {Ferrighi}, \citenamefont {Gavnholt}, \citenamefont {Glinsvad}, \citenamefont {Haikola}, \citenamefont {Hansen} \emph {et~al.}}]{enkovaara2010electronic}%
  \BibitemOpen
  \bibfield  {author} {\bibinfo {author} {\bibfnamefont {J.}~\bibnamefont {Enkovaara}}, \bibinfo {author} {\bibfnamefont {C.}~\bibnamefont {Rostgaard}}, \bibinfo {author} {\bibfnamefont {J.~J.}\ \bibnamefont {Mortensen}}, \bibinfo {author} {\bibfnamefont {J.}~\bibnamefont {Chen}}, \bibinfo {author} {\bibfnamefont {M.}~\bibnamefont {Du{\l}ak}}, \bibinfo {author} {\bibfnamefont {L.}~\bibnamefont {Ferrighi}}, \bibinfo {author} {\bibfnamefont {J.}~\bibnamefont {Gavnholt}}, \bibinfo {author} {\bibfnamefont {C.}~\bibnamefont {Glinsvad}}, \bibinfo {author} {\bibfnamefont {V.}~\bibnamefont {Haikola}}, \bibinfo {author} {\bibfnamefont {H.}~\bibnamefont {Hansen}},  \emph {et~al.},\ }\href@noop {} {\bibfield  {journal} {\bibinfo  {journal} {Journal of physics: Condensed matter}\ }\textbf {\bibinfo {volume} {22}},\ \bibinfo {pages} {253202} (\bibinfo {year} {2010})}\BibitemShut {NoStop}%
\bibitem [{\citenamefont {Torelli}\ and\ \citenamefont {Olsen}(2018)}]{torelli2018calculating}%
  \BibitemOpen
  \bibfield  {author} {\bibinfo {author} {\bibfnamefont {D.}~\bibnamefont {Torelli}}\ and\ \bibinfo {author} {\bibfnamefont {T.}~\bibnamefont {Olsen}},\ }\href@noop {} {\bibfield  {journal} {\bibinfo  {journal} {2D Materials}\ }\textbf {\bibinfo {volume} {6}},\ \bibinfo {pages} {015028} (\bibinfo {year} {2018})}\BibitemShut {NoStop}%
\bibitem [{\citenamefont {Olsen}(2019)}]{Olsen2019}%
  \BibitemOpen
  \bibfield  {author} {\bibinfo {author} {\bibfnamefont {T.}~\bibnamefont {Olsen}},\ }\href {\doibase 10.1557/mrc.2019.117} {\bibfield  {journal} {\bibinfo  {journal} {MRS Communications}\ }\textbf {\bibinfo {volume} {9}},\ \bibinfo {pages} {1142} (\bibinfo {year} {2019})},\ \Eprint {http://arxiv.org/abs/1908.11115} {arXiv:1908.11115} \BibitemShut {NoStop}%
\bibitem [{\citenamefont {Pakdel}\ \emph {et~al.}(2023)\citenamefont {Pakdel}, \citenamefont {Rasmussen}, \citenamefont {Taghizadeh}, \citenamefont {Kruse}, \citenamefont {Olsen},\ and\ \citenamefont {Thygesen}}]{pakdel2023emergent}%
  \BibitemOpen
  \bibfield  {author} {\bibinfo {author} {\bibfnamefont {S.}~\bibnamefont {Pakdel}}, \bibinfo {author} {\bibfnamefont {A.}~\bibnamefont {Rasmussen}}, \bibinfo {author} {\bibfnamefont {A.}~\bibnamefont {Taghizadeh}}, \bibinfo {author} {\bibfnamefont {M.}~\bibnamefont {Kruse}}, \bibinfo {author} {\bibfnamefont {T.}~\bibnamefont {Olsen}}, \ and\ \bibinfo {author} {\bibfnamefont {K.~S.}\ \bibnamefont {Thygesen}},\ }\href@noop {} {\bibfield  {journal} {\bibinfo  {journal} {arXiv preprint arXiv:2304.01148}\ } (\bibinfo {year} {2023})}\BibitemShut {NoStop}%
\bibitem [{\citenamefont {Huang}\ \emph {et~al.}(2017)\citenamefont {Huang}, \citenamefont {Clark}, \citenamefont {Navarro-Moratalla}, \citenamefont {Klein}, \citenamefont {Cheng}, \citenamefont {Seyler}, \citenamefont {Zhong}, \citenamefont {Schmidgall}, \citenamefont {McGuire}, \citenamefont {Cobden} \emph {et~al.}}]{huang2017layer}%
  \BibitemOpen
  \bibfield  {author} {\bibinfo {author} {\bibfnamefont {B.}~\bibnamefont {Huang}}, \bibinfo {author} {\bibfnamefont {G.}~\bibnamefont {Clark}}, \bibinfo {author} {\bibfnamefont {E.}~\bibnamefont {Navarro-Moratalla}}, \bibinfo {author} {\bibfnamefont {D.~R.}\ \bibnamefont {Klein}}, \bibinfo {author} {\bibfnamefont {R.}~\bibnamefont {Cheng}}, \bibinfo {author} {\bibfnamefont {K.~L.}\ \bibnamefont {Seyler}}, \bibinfo {author} {\bibfnamefont {D.}~\bibnamefont {Zhong}}, \bibinfo {author} {\bibfnamefont {E.}~\bibnamefont {Schmidgall}}, \bibinfo {author} {\bibfnamefont {M.~A.}\ \bibnamefont {McGuire}}, \bibinfo {author} {\bibfnamefont {D.~H.}\ \bibnamefont {Cobden}},  \emph {et~al.},\ }\href@noop {} {\bibfield  {journal} {\bibinfo  {journal} {Nature}\ }\textbf {\bibinfo {volume} {546}},\ \bibinfo {pages} {270} (\bibinfo {year} {2017})}\BibitemShut {NoStop}%
\bibitem [{\citenamefont {Wilson}\ \emph {et~al.}(2021)\citenamefont {Wilson}, \citenamefont {Lee}, \citenamefont {Cenker}, \citenamefont {Xie}, \citenamefont {Dismukes}, \citenamefont {Telford}, \citenamefont {Fonseca}, \citenamefont {Sivakumar}, \citenamefont {Dean}, \citenamefont {Cao} \emph {et~al.}}]{wilson2021interlayer}%
  \BibitemOpen
  \bibfield  {author} {\bibinfo {author} {\bibfnamefont {N.~P.}\ \bibnamefont {Wilson}}, \bibinfo {author} {\bibfnamefont {K.}~\bibnamefont {Lee}}, \bibinfo {author} {\bibfnamefont {J.}~\bibnamefont {Cenker}}, \bibinfo {author} {\bibfnamefont {K.}~\bibnamefont {Xie}}, \bibinfo {author} {\bibfnamefont {A.~H.}\ \bibnamefont {Dismukes}}, \bibinfo {author} {\bibfnamefont {E.~J.}\ \bibnamefont {Telford}}, \bibinfo {author} {\bibfnamefont {J.}~\bibnamefont {Fonseca}}, \bibinfo {author} {\bibfnamefont {S.}~\bibnamefont {Sivakumar}}, \bibinfo {author} {\bibfnamefont {C.}~\bibnamefont {Dean}}, \bibinfo {author} {\bibfnamefont {T.}~\bibnamefont {Cao}},  \emph {et~al.},\ }\href@noop {} {\bibfield  {journal} {\bibinfo  {journal} {Nature Materials}\ }\textbf {\bibinfo {volume} {20}},\ \bibinfo {pages} {1657} (\bibinfo {year} {2021})}\BibitemShut {NoStop}%
\bibitem [{\citenamefont {Kong}\ \emph {et~al.}(2019)\citenamefont {Kong}, \citenamefont {Stolze}, \citenamefont {Timmons}, \citenamefont {Tao}, \citenamefont {Ni}, \citenamefont {Guo}, \citenamefont {Yang}, \citenamefont {Prozorov},\ and\ \citenamefont {Cava}}]{kong2019vi3}%
  \BibitemOpen
  \bibfield  {author} {\bibinfo {author} {\bibfnamefont {T.}~\bibnamefont {Kong}}, \bibinfo {author} {\bibfnamefont {K.}~\bibnamefont {Stolze}}, \bibinfo {author} {\bibfnamefont {E.~I.}\ \bibnamefont {Timmons}}, \bibinfo {author} {\bibfnamefont {J.}~\bibnamefont {Tao}}, \bibinfo {author} {\bibfnamefont {D.}~\bibnamefont {Ni}}, \bibinfo {author} {\bibfnamefont {S.}~\bibnamefont {Guo}}, \bibinfo {author} {\bibfnamefont {Z.}~\bibnamefont {Yang}}, \bibinfo {author} {\bibfnamefont {R.}~\bibnamefont {Prozorov}}, \ and\ \bibinfo {author} {\bibfnamefont {R.~J.}\ \bibnamefont {Cava}},\ }\href@noop {} {\bibfield  {journal} {\bibinfo  {journal} {Advanced Materials}\ }\textbf {\bibinfo {volume} {31}},\ \bibinfo {pages} {1808074} (\bibinfo {year} {2019})}\BibitemShut {NoStop}%
\bibitem [{\citenamefont {Hovancik}\ \emph {et~al.}(2023)\citenamefont {Hovancik}, \citenamefont {Pospisil}, \citenamefont {Carva}, \citenamefont {Sechovsky},\ and\ \citenamefont {Piamonteze}}]{hovancik2023large}%
  \BibitemOpen
  \bibfield  {author} {\bibinfo {author} {\bibfnamefont {D.}~\bibnamefont {Hovancik}}, \bibinfo {author} {\bibfnamefont {J.}~\bibnamefont {Pospisil}}, \bibinfo {author} {\bibfnamefont {K.}~\bibnamefont {Carva}}, \bibinfo {author} {\bibfnamefont {V.}~\bibnamefont {Sechovsky}}, \ and\ \bibinfo {author} {\bibfnamefont {C.}~\bibnamefont {Piamonteze}},\ }\href@noop {} {\bibfield  {journal} {\bibinfo  {journal} {Nano Letters}\ }\textbf {\bibinfo {volume} {23}},\ \bibinfo {pages} {1175} (\bibinfo {year} {2023})}\BibitemShut {NoStop}%
\bibitem [{\citenamefont {Olsen}(2016)}]{Olsen2016a}%
  \BibitemOpen
  \bibfield  {author} {\bibinfo {author} {\bibfnamefont {T.}~\bibnamefont {Olsen}},\ }\href {\doibase 10.1103/PhysRevB.94.235106} {\bibfield  {journal} {\bibinfo  {journal} {Physical Review B}\ }\textbf {\bibinfo {volume} {94}},\ \bibinfo {pages} {235106} (\bibinfo {year} {2016})}\BibitemShut {NoStop}%
\bibitem [{\citenamefont {SÃ¸dequist}\ and\ \citenamefont {Olsen}(2023)}]{sÃ¸dequist2023magnetic}%
  \BibitemOpen
  \bibfield  {author} {\bibinfo {author} {\bibfnamefont {J.}~\bibnamefont {SÃ¸dequist}}\ and\ \bibinfo {author} {\bibfnamefont {T.}~\bibnamefont {Olsen}},\ }\href@noop {} {\enquote {\bibinfo {title} {Magnetic order in the computational 2d materials database (c2db) from high throughput spin spiral calculations},}\ } (\bibinfo {year} {2023}),\ \Eprint {http://arxiv.org/abs/2309.11945} {arXiv:2309.11945 [cond-mat.mtrl-sci]} \BibitemShut {NoStop}%
\end{thebibliography}

%

\end{document}